\documentclass[sigconf]{acmart}

\usepackage{comment}

\usepackage{booktabs} \usepackage{multibib}
\newcites{latex}{Additional References}

\usepackage{mdwlist}

\setcopyright{rightsretained}

\acmConference{KDD 2019}{Aug 4--8}{Anchorage}
\acmYear{2019}
\copyrightyear{2019}

\usepackage{enumitem}

\newtheorem{Lemma}{Lemma}[section]  
\newtheorem{Definition}{Definition}

\usepackage{footmisc}
\usepackage{comment}
\usepackage{tabularx}
\usepackage{mathrsfs}
\usepackage{amsmath}
\usepackage{amssymb}
\usepackage{algorithm}
\usepackage{tikz}
\usepackage{xspace}
\usepackage{graphicx}
\usepackage{multirow}
\usepackage{subcaption}
\usepackage{amssymb}\usepackage{pifont}\newcommand{\cmark}{\ding{51}}\newcommand{\xmark}{\ding{55}}\usetikzlibrary{matrix,decorations.pathreplacing,calc}
\usepackage{mathtools}

\pgfkeys{tikz/mymatrixenv/.style={decoration=brace,every left delimiter/.style={xshift=3pt},every right delimiter/.style={xshift=-3pt}}}
\pgfkeys{tikz/mymatrix/.style={matrix of math nodes,left delimiter=[,right delimiter={]},inner sep=2pt,column sep=1em,row sep=0.5em,nodes={inner sep=0pt}}}
\pgfkeys{tikz/mymatrixbrace/.style={decorate,thick}}

\newcommand{\Phisize}{|\Phi|}

\newcommand{\B}{\mathcal{B}}
\newcommand{\Bsize}{|\mathcal{B}|}
\newcommand{\T}{\mathcal{T}}

\newcommand{\F}{\mathcal{F}}

\newcommand{\Fr}{\mathcal{F}_r}

\newcommand{\hist}[1]{\mathbf{h}_{#1}}

\usepackage{xcolor}
\definecolor{thedarkblue}{RGB}{0,0,120} \definecolor{mydarkblue}{rgb}{0,0.08,0.45} 
\usepackage{hyperref}
\hypersetup{    colorlinks=true,
    linkcolor=black,
    citecolor=black,
    filecolor=mydarkblue,
    urlcolor=mydarkblue}
    
\usepackage{microtype}
\usepackage{balance}

\newcolumntype{L}[1]{>{\raggedright\let\newline\\\arraybackslash\hspace{0pt}}m{#1}}
\newcolumntype{C}[1]{>{\centering\let\newline\\\arraybackslash\hspace{0pt}}m{#1}}
\newcolumntype{R}[1]{>{\raggedleft\let\newline\\\arraybackslash\hspace{0pt}}m{#1}}

\providecommand{\mat}[1]{\boldsymbol{\mathrm{#1}}}\renewcommand{\vec}[1]{\boldsymbol{\mathrm{#1}}}

\DeclareMathOperator{\hugeE}{\mbox{\huge\raise-0.3ex\hbox{E}}}
\DeclareMathOperator{\p}{\mathbb{P}}
\DeclareMathOperator{\hugep}{\mbox{\huge\raise-0.3ex\hbox{$\p$}}}

\newcommand{\method}{\textsc{Multi-LENS}\xspace}

\providecommand{\mA}{\ensuremath{\mat{A}}}

\providecommand{\mH}{\ensuremath{\mat{S}}}

\providecommand{\mU}{\ensuremath{\mat{U}}}
\providecommand{\mV}{\ensuremath{\mat{V}}}

\providecommand{\mX}{\ensuremath{\mat{F}}}
\providecommand{\mXl}{\ensuremath{\mat{F}^{(\ell)}}}
\providecommand{\mXo}{\ensuremath{\mat{F}^{(0)}}}
\providecommand{\mY}{\ensuremath{\mat{H}}}

\providecommand{\vb}{\ensuremath{\vec{b}}}

\providecommand{\vx}{\ensuremath{\vec{x}}}

\usepackage{setspace}
\graphicspath{{./}{./graphics/}}
\newcolumntype{H}{>{\setbox0=\hbox\bgroup}c<{\egroup}@{}}

\newcommand{\eg}{\emph{e.g.}}
\newcommand{\ie}{\emph{i.e.}}

\newcommand\TT{\rule{0pt}{2.7ex}}
\newcommand\BB{\rule[-1.2ex]{0pt}{0pt}}
\newcommand{\neighborhood}{\mathcal{N}}

\usepackage{tikz}
\usepackage{verbatim}
\usetikzlibrary{arrows}
\usetikzlibrary{shapes,snakes}
\usetikzlibrary{decorations.pathmorphing} \usetikzlibrary{fit}					\usetikzlibrary{backgrounds}	
\usepackage{algorithm}
\usepackage{algorithmicx}
\usepackage{algpseudocode}

\algblockdefx[parallel]{ParFor}{EndPar}[1][]{$\textbf{parallel for}$ #1 $\textbf{do}$}{$\textbf{end parallel}$}
\algrenewcommand{\alglinenumber}[1]{\fontsize{6.5}{7}\selectfont#1}
\algtext*{EndPar}

\algblockdefx[parallel]{parfor}{endpar}[1][]{$\textbf{parallel for}$ #1 $\textbf{do}$}{$\textbf{end parallel}$}
\algrenewcommand{\alglinenumber}[1]{\scriptsize#1:}

\algblockdefx[foreach1]{foreach}{endforeach}[1][]{$\textbf{for}$ #1 $\textbf{do}$}{$\textbf{end for}$}
\algrenewcommand{\alglinenumber}[1]{\scriptsize#1:}

\algnotext{EndFor}
\algnotext{EndIf}
\algnotext{EndProcedure}



\algblockdefx[fornew1]{foreach}{endforeach}
[1][]{\textbf{for} #1}
{\textbf{end for}}

\begin{document}
\title{Latent Network Summarization: Bridging Network Embedding and Summarization}

\author{Di Jin}
\affiliation{  \institution{University of Michigan}
}
\email{dijin@umich.edu}

\author{Ryan A. Rossi}
\orcid{1234-5678-9012-3456}
\affiliation{  \institution{Adobe Research}
}
\email{rrossi@adobe.com}

\author{Eunyee Koh}
\affiliation{  \institution{Adobe Research}
}
\email{eunyee@adobe.com}

\author{Sungchul Kim}
\affiliation{  \institution{Adobe Research}
}
\email{sukim@adobe.com}
\email{}
\author{Anup Rao}
\affiliation{  \institution{Adobe Research}
}
\email{anuprao@adobe.com}

\author{Danai Koutra}
\affiliation{  \institution{University of Michigan}
}
\email{dkoutra@umich.edu}

\renewcommand\shortauthors{Jin, D. et al}

\begin{abstract}
Motivated by the computational and storage challenges that dense embeddings pose, we introduce the problem of \emph{latent network summarization} that aims to learn a compact, latent representation of the graph structure with dimensionality that is \textit{independent} of the input graph size (\ie, \#nodes and \#edges), while retaining the ability to derive node representations on the fly. 
We propose \method, 
an inductive multi-level latent network summarization approach that leverages a set of \emph{relational operators} and \textit{relational functions} (compositions of operators) to capture the structure of egonets and higher-order subgraphs, respectively. The structure is stored in low-rank, size-independent structural feature matrices, which along with the relational functions comprise our latent network summary.  
\method is general and naturally supports both homogeneous \emph{and} heterogeneous graphs with or without directionality, weights, attributes or labels.
Extensive experiments on real graphs 
show $3.5-34.3\%$ improvement in AUC for link prediction, 
while requiring $80-2152\times$ \textit{less} output storage space than baseline embedding methods on \textit{large} datasets. 
As application areas, we show the effectiveness of \method in detecting anomalies and events in the Enron email communication graph and Twitter co-mention graph.

\end{abstract}

\copyrightyear{2019} 
\acmYear{2019} 
\setcopyright{acmcopyright}
\acmConference[KDD '19]{The 25th ACM SIGKDD Conference on Knowledge Discovery and Data Mining}{August 4--8, 2019}{Anchorage, AK, USA}
\acmBooktitle{The 25th ACM SIGKDD Conference on Knowledge Discovery and Data Mining (KDD '19), August 4--8, 2019, Anchorage, AK, USA}
\acmPrice{15.00}
\acmDOI{10.1145/3292500.3330992}
\acmISBN{978-1-4503-6201-6/19/08}

\settopmatter{printacmref=true}
\fancyhead{}

\maketitle

\section{Introduction}
\label{sec_intro}

\begin{figure}[t]
	\centering
	\includegraphics[width=0.9\linewidth]{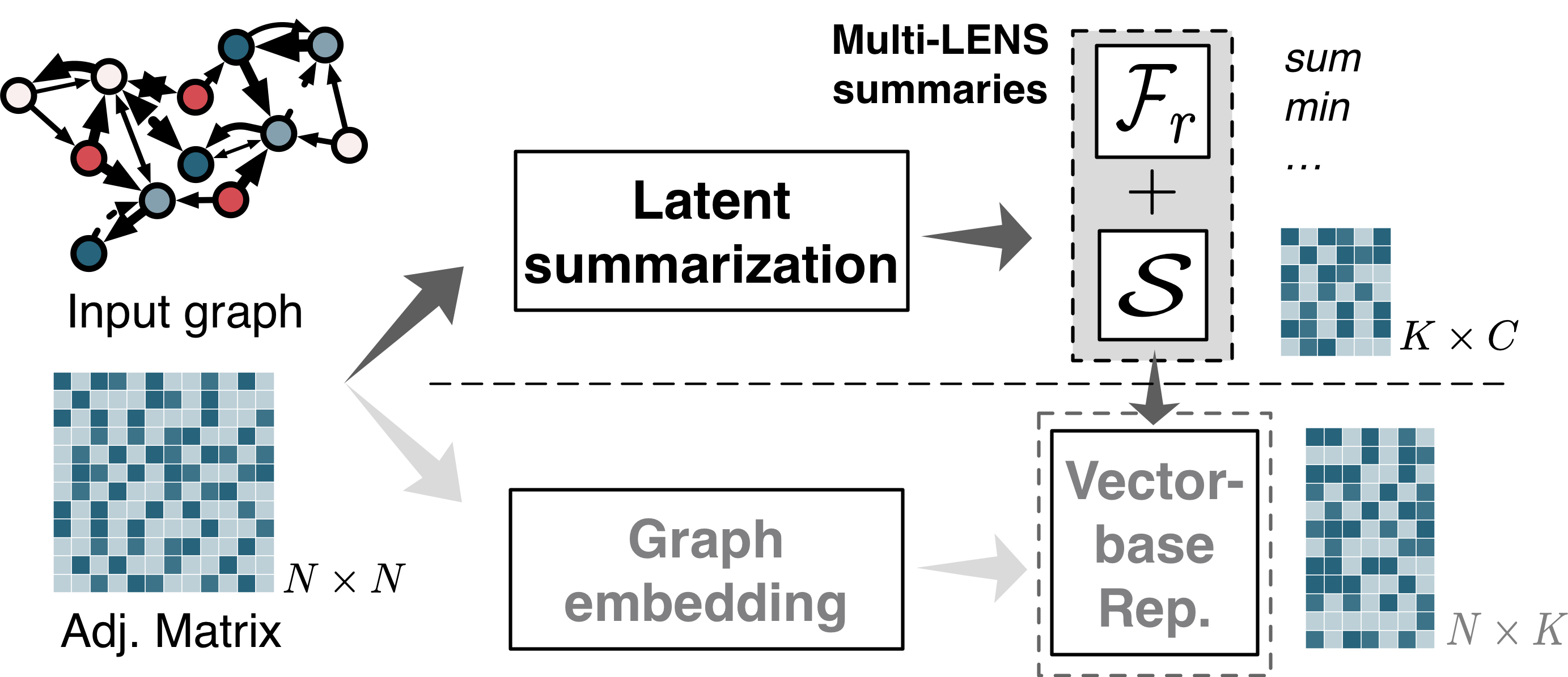}\vspace{-0.25cm}
	\caption{Our proposed approach to Latent Network Summarization called \method produces a summary consisting of relational functions $\Fr$ and 
		node-independent matrices $\mathcal{S}$ of size $K\times C$. 
		Thus, while embedding methods output $N$ node embeddings of dimensionality $K$, latent summarization methods produce an output that is independent of $N$ and thus is graph-size independent.  
		Despite not storing the embeddings, \method can derive them on the fly.
	}
	\label{fig_highlight}
	\vspace{-0.6 cm}
\end{figure}

\enlargethispage{\baselineskip}

Recent advances in representation learning for graphs have led to a variety of proximity-based and structural embeddings that achieve superior performance in specific downstream tasks, such as link prediction, node classification, and alignment~\cite{goyal2018graph,deepGL,regal}. At the same time though, the learned, $K$-dimensional node embeddings are dense (with real values), and pose computational and storage challenges especially for massive graphs. By following the conventional setting of $K=128$ for the dimensionality, a graph of 1 billion nodes requires roughly 1TB for its embeddings. Moreover, this dense representation often requires significantly more space to store than the original, sparse adjacency matrix of a graph. For example, for the datasets that we consider in our empirical analysis, the learned embeddings from existing representation learning techniques require $3-48\times$ more space than the original edge files.

To address these shortcomings, we introduce the problem of \textit{latent network summarization}.
Informally, the goal is to find a low-dimensional representation in a latent space   such that it is \emph{independent} of the graph size, \ie, the number of nodes and edges. 
Among other tasks, the representation should support \emph{on-the-fly} computation of \emph{specific} node embeddings.
Latent network summarization and network embedding are \emph{complementary} learning tasks with fundamentally different goals and outputs, as shown in Fig.~\ref{fig_highlight}.
In particular, the goal of network embedding is to derive $N$ 
node embedding vectors of $K$ dimensions each that capture node proximity or equivalency. 
Thus, the output is a $N\times K$ matrix that is \emph{dependent} on the size of the graph (number of nodes)~\cite{qiu2018network,goyal2018graph}.
This is in contrast to the goal of latent network summarization, which is to learn a \emph{size-independent representation} of the graph.
\textit{Latent} network summarization also differs from traditional summarization approaches that typically derive supergraphs (\eg, mapping nodes to supernodes)~\cite{liu2018graph}, which target different applications and are unable to derive node embeddings.

To efficiently solve the latent network summarization problem, we propose \method (Multi-level Latent Network Summarization), an inductive framework that is based on graph function compositions.
In a nutshell, the method begins with a set of arbitrary graph features (\eg, degree) and iteratively uses generally-defined relational operators over neighborhoods to derive deeper function compositions that capture graph features at multiple \textit{levels} (or distances).
Low-rank approximation is then used to derive the best-fit subspace vectors of network features across levels. 
Thus, the latent summary given by \method comprises graph functions and latent vectors, both of which are independent of the graph size.
Our main contributions are summarized as follows:

\begin{itemize}[leftmargin=12pt]
	\item  \textbf{Novel Problem Formulation.} 
	We introduce and formulate the problem of \emph{latent network summarization}, which is complementary yet fundamentally different from network embedding. 
	
	\item  \textbf{Computational Framework.} We propose \method, which expresses a class of methods for latent network summarization. \method naturally supports inductive learning, on-the-fly embedding computation for all or a \textit{subset} of nodes.

	\item   \textbf{Time- and Space-efficiency.} 
	\method is \textit{scalable} with time complexity linear on the number of edges, and \textit{space-efficient} with size independent of the graph size. Besides, it is \textit{parallelizable} as the node computations are independent of each other.
	
	\item  \textbf{Empirical analysis on real datasets.}
	We apply \method 
	to event detection and link prediction over real-world heterogeneous graphs and show that it is $3.5\%$-$34.3\%$ more accurate than state-of-the-art embedding methods while requiring $80$-$2152\times$ less output storage space for datasets with \textit{millions} of edges.
\end{itemize}

Next we formally introduce the latent network summarization problem and then describe our proposed framework.

\section{Latent Network Summarization}
\label{sec_latent_graph_summarization}
\providecommand{\mPhi}{\ensuremath{{\rm \boldsymbol\Phi}}}

Intuitively, the problem of \textit{latent network summarization} aims to learn a compressed representation that captures the main structural information of the network and depends only on the complexity of the network instead of its size. More formally:

\begin{Definition}[Latent Network Summarization] \label{def_latent_graph_summarization}
	Given an arbitrary graph $G = (V, E)$ 
	with $|V|=N$ nodes and $|E|=M$ edges, 
	the goal of latent network summarization is to map the graph $G$ to a low-dimensional $K \times C$ representation $\mathcal{J}$ that summarizes the structure of $G$, where  $K, C \ll N, M$ are independent of the graph size. 
	The output latent representations should be usable in data mining tasks, and sufficient to derive all or a subset of node embeddings on the fly for learning tasks (e.g., link prediction, classification). \end{Definition}

Compared to the network embedding problem, 
latent network summarization differs in that it aims to 
derive a \textit{size-independent} representation of the graph.
This can be achieved in the form of supergraphs~\cite{liu2018graph} (in the original graph space) or aggregated clusters trivially, but the compressed latent network summary in Definition~\ref{def_latent_graph_summarization} also needs to be able to derive the node embeddings, which is not the goal of traditional graph summarization methods. 

In general, based on our definition, a latent network summarization approach should satisfy the following key properties: {\bf(P1)} generality to handle arbitrary network with multiple node types, relationship types, edge weights, directionality, unipartite or k-partite structure, \emph{etc.}
{\bf(P2)} high compression rate, {\bf(P3)} natural support of inductive learning, and {\bf(P4)}~ability to on-the-fly derive node embeddings used in follow-up tasks.

\begin{table}[t!]
	\centering
	\caption{Summary of symbols and notations}
	\vspace{-3mm}
	\centering 
	\fontsize{7}{7.5}\selectfont
	\setlength{\tabcolsep}{5pt} \label{table_symbols}
	\vspace{-0.1cm}
	\def\arraystretch{1.25} \begin{tabularx}{1.0\linewidth}{@{}rX@{}}
		\toprule
		\textbf{Symbol} & \textbf{Definition} 
		\\ 
		\midrule
		
		$G=(V, E)$ & heterogeneous network with $|V|=N$ nodes and $|E|=M$ edges  \\
		$\mA$ &  adjacency matrix of $G$ with row $i$ $\mA_{i,:}$ and column $i$ $\mA_{:,i}$ \\
		$\T_V, \T_E$ & sets of object types and edge types, respectively \\
		$\neighborhood_i, \neighborhood^{t}_i$ & non-typed / types (1-hop) neighborhood or egonet of node $i$ \\
		$\ell$, $L$ & index for level \& total number of levels (\ie, max order of a rel.\ fns) \\
		$\B$ & =$\{\vb_i\}$ set of initial feature vectors in length $N$\\
		$\Fr$ & =$\{\Fr^{(1)},\dots,\Fr^{(L)}\}$,  ordered set of relational functions across levels \\ $\F_b$ & = $\{f_{b_i}\}$, set of base graph functions (special relational functions) \\
		$\Phi$ &  = $\{\phi_{i}\}$, set of relational operators \\ 
		
		$\mX^{(0)}$  & $N\times \Bsize$ base feature matrix derived by the base graph functions $\mathcal{F}_b$\\
		$\mX^{(\ell)}$ & $N\times \left(\Bsize \cdot \Phisize^{\ell} \right)$ generated feature  matrix for level $\ell$ \\

		$K^{(\ell)}$, $K$ & dimensionality of embeddings at level-$\ell$ and the final dimensionality\\
		
		$\mY^{(\ell)}$ & $N\times |\Fr^{(\ell)}|$ histogram-based representation of feature matrix $\mX^{(\ell)}$  \\
		$\mH^{(\ell)}$ & low-rank latent graph summary at level $\ell$ \\
		\bottomrule
	\end{tabularx}
	\vspace{-0.5cm}
\end{table}

\section{\method Framework} \label{sec_approach}

\begin{figure}[th!]
	\centering
	\includegraphics[width=1.0\linewidth]{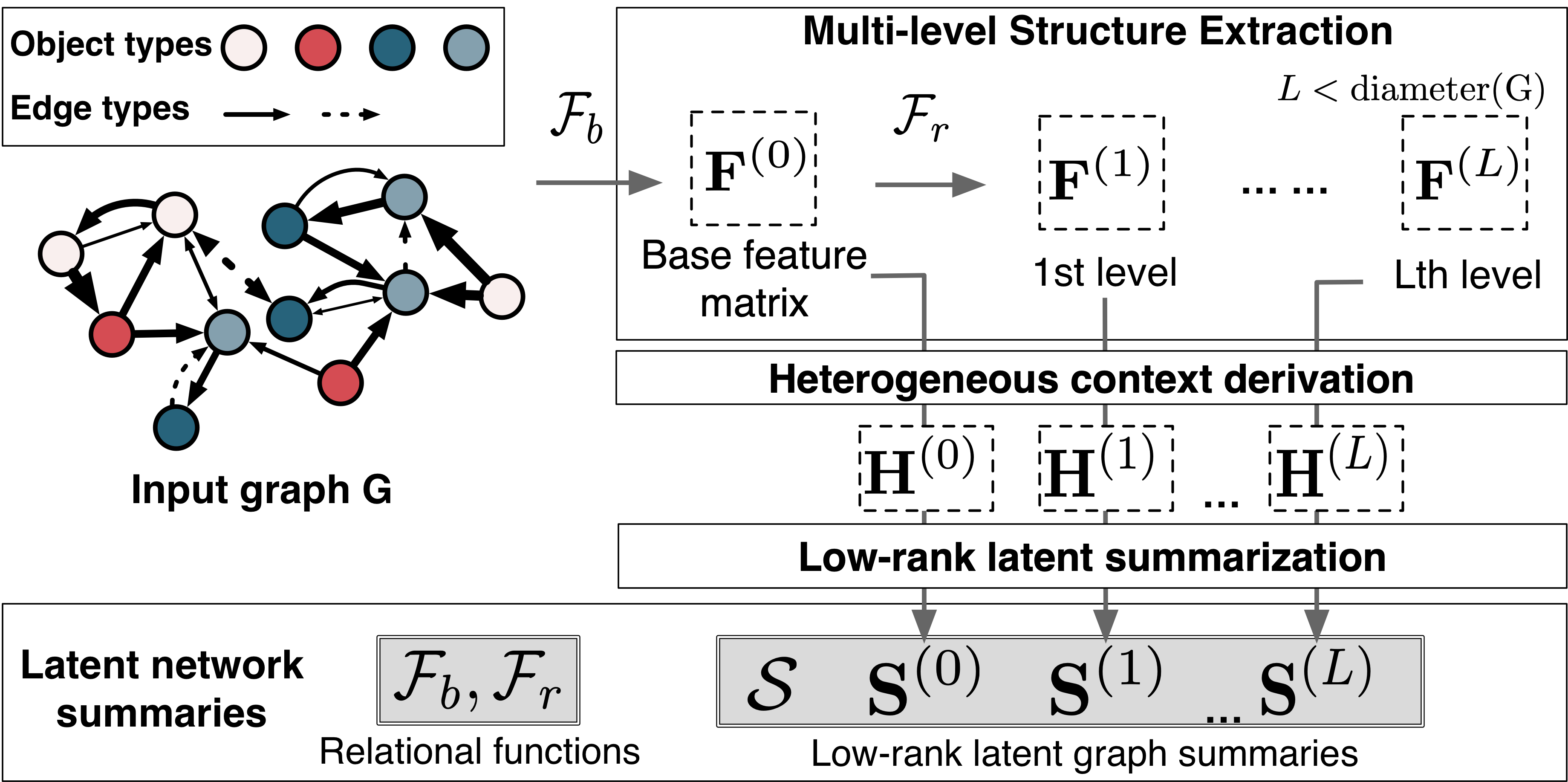}
	\vspace{-0.5cm}
	\caption{Overview of \method. Dashed boxes: intermediate results that do not need to store; shaded boxes: outputs that need storing. The size of the latent network summaries, $\mathcal{J}=\big\{\mathcal{F}, \mathcal{S}\big\}$, is independent of $N, M$.
	}
	\label{fig_workflow}
	\vspace{-0.4cm}
\end{figure}

To efficiently address the problem of latent network summarization introduced in Section~\ref{sec_latent_graph_summarization}, we propose \method, which expresses a class of latent network summarization methods that satisfies all desired properties {\bf (P1-P4)}. 
The summary $\mathcal{J}$ given by \method contains (i) necessary operators for aggregating node-wise structural features automatically and (ii) subspace vectors on which to derive the embeddings.
We give the overview in Figure~\ref{fig_workflow} and list the main symbols and notations used in this work in Table~\ref{table_symbols}.

At a high level, \method leverages generally-defined relational operators to capture structural information from node neighborhoods in arbitrary types of networks.
It recursively applies these operators over node neighborhoods to produce both linear and non-linear functions that characterize each node at different distances (\S~\ref{sec_multi_level_structure_extraction}).
To efficiently derive the contextual space vectors, \method first generates histogram-based heterogeneous contexts for nodes (\S~\ref{sec_summarizing_hetero}), and then obtains the summary via low-dimensional approximation (\S~\ref{sec_summarization}).
We include the empirical justification of our design choices in the Appendix.
Before discussing each step and its rationale, we first present some preliminaries that serve as building blocks for \method.

\vspace{-0.1cm}
\subsection{Preliminaries}
\label{sec_handling_arbitrary_networks}

Recall that our proposed problem definition (\S~\ref{sec_latent_graph_summarization}) applies to any arbitrary graph {\bf (P1)}. 
As a general class, we refer to heterogeneous (information) networks or typed networks.  \begin{Definition}[Heterogeneous network] \label{def:heter-network}
	A heterogeneous network is defined as $G=(V,E,\theta,\xi)$ with node-set $V$, edge-set $E$,   a function $\,\theta : V \rightarrow \mathcal{T}_V$ mapping nodes to their types,  
	and a function $\,\xi : E \rightarrow \mathcal{T}_E$ mapping edges to their types.  
\end{Definition}\noindent
We assume that the network is directed and weighted with unweighted and undirected graphs as special cases. For simplicity, we will refer to a graph as $G(V, E)$. 
Within heterogeneous networks, the typed neighborhood or egonet\footnote{In this work we use neighborhood and egonet interchangeably.} $\neighborhood_{t}$ of a node is defined as follows:
\begin{Definition}[Typed neighborhood $\neighborhood^{t}$]\label{def:typed-l-neighborhood}
	Given an arbitrary node $i$ in graph $G=(V, E)$, the typed $t$ \emph{neighborhood} $\neighborhood^{t}_i$ is the set of nodes with type $t$ that are reachable by following directed edges $e\in E $ originating from $i$ with $1$-hop distance and $i$ itself.
\end{Definition}\noindent
The \emph{neighborhood} of node $i$, $\neighborhood_i$, is a superset of the typed neighborhood $\neighborhood^{t}_i$, and includes nodes in the neighborhood of $i$ regardless of their types. 
Higher-order neighborhoods are defined similarly, but more computationally expensive to explore. For example, the $k$-hop neighborhood denotes the set of nodes reachable following directed edges $e\in E $ originating from node $i$ within $k$-hop distance. 

The goal of latent network summarization  is to find a size-independent representation that captures the \textit{structure} of the network and its underlying nodes in the latent space.
Capturing the structure depends on the semantics of the network (e.g., weighted, directed), and thus different ways are needed for different input networks types. 
To generalize to \textit{arbitrary} networks, we leverage relational operators and functions~\cite{deepGL}.

\begin{Definition}[\sc Relational operator] \label{def:relational-operator}
	A relational operator $\phi(\vx, \mathcal{R})\in\Phi$ is defined as a basic function (\eg, sum) that operates on a feature vector $\vx$ associated with a set of related elements $\mathcal{R}$ and returns a single value.
\end{Definition}\noindent

For example, let $\vx$ be an $N \times 1$ vector and $\mathcal{R}$ the neighborhood $\neighborhood_i$ of node $i$. For $\phi$ being the \emph{sum}, $\phi(\vx, \mathcal{R})$ would return the count of neighbors reachable from node $i$ (unweighted out degree).

\begin{Definition}[\sc Relational function] \label{def:relational-function}
	\vspace{-0.1cm}
	A relational function $f \in \mathcal{F}$ is defined as a composition of relational operators $f=\big( \, \phi_1 \circ \cdots \circ \phi_{h-1} \circ \phi_h\big)(\vx, \mathcal{R})$ applied to feature values in $\mathbf{x}$ associated with the set of related nodes $\mathcal{R}$. We say that $f$ is order-$h$ iff the feature vector $\vx$ is applied to $h$ relational operators. \end{Definition}\noindent

Together, relational operators and relational functions comprise the building blocks of our proposed method, \method.
Iterative computations over the graph or a subgraph (\eg, node neighborhood) generalize for inductive/across-network transfer learning tasks. 
Moreover, relational functions are general and can be used to derive commonly-used graph statistics.
As an example, the out-degree of a specific node is derived by applying order-1 relational functions on the adjacency matrix over its the egonet, \ie, $\text{out-deg(i)} = \sum(\mathbf{A}_{i:}, \neighborhood)$ regardless of object types.

\vspace{-0.1cm}
\subsection{Multi-level Structure Extraction}
\label{sec_multi_level_structure_extraction}
We now start describing our proposed method, \method. The first step is to extract multi-level strcuture around the nodes. To this end, as we show in Figure~\ref{fig_workflow}, \method first generates a set of simple node-level features to form the base feature matrix $\mX^{(0)}$ via the so-called base graph functions $\mathcal{F}_b$.
It then composes new functions by iteratively applying a set of relational operators $\Phi$  over the neighborhood to generate new features. 
Operations in both $\mathcal{F}_b$ and $\Phi$ are generally defined to satisfy ${\bf (P1)}$.

\subsubsection{Base Graph Functions}\label{sec:base-functions}
As a special relational function, each base graph function $f_b\in\mathcal{F}_b$ consists of relational operators that perform on an initial $N\times 1$ feature vector $\vb\in\mathcal{B}$. 
The vector $\vb$ could be given as the row/column of the adjacency matrix corresponding to node $i$, or some other derived vector related to the node (\eg, its distance or influence to every node in the graph).
Following~\cite{deepGL}, the simplest case is $f_b=\sum$, which captures simple base features such as in/out/total degrees.
We denote applying the same base function to the egonets of all the nodes in graph $G$ as follows:
\begin{equation}
f_b\langle\vb, \neighborhood\rangle =[f_b(\vb, \neighborhood_1), f_b(\vb, \neighborhood_2), \dots, f_b(\vb, \neighborhood_N)]^{T}, \vb\in \mathcal{B} \vspace{1cm}
\label{eq_base_vector}
\vspace{-1cm}
\end{equation}
which forms an $N\times 1$ vector. For example, $f_b=\sum\langle \mathbf{A}_{i:}, \neighborhood \rangle$ enumerates the out-degree of all nodes in $G$.
By applying $f_b$ on each initial feature $\vb$, \eg, $\mathbf{1}^{N\times 1}$ or row/column of adjacency matrix $\mA$, we obtain the $N\times B$ base matrix $\mX^{(0)}$:
\begin{equation}
\mX^{(0)} = [f_b\langle\vb_1, \neighborhood\rangle, f_b\langle\vb_2, \neighborhood\rangle, \dots, f_b\langle\vb_B, \neighborhood\rangle], \vb_{1\cdots B} \in \mathcal{B}
\label{eq_x0}
\end{equation}
which aggregates all structural features of the nodes within $\neighborhood$.
The specific choice of initial vectors $\vb$ is not very important as the composed relational functions (\S~\ref{sec_function_composition}) extensively incorporate both linear and nonlinear structural information automatically. We empirically justify \method on the link prediction task over different choices of $\mathcal{B}$ to show its insensitivity in Appendix~\S~\ref{sec_justify_init_features}.

\subsubsection{Relational Function Compositions} \label{sec_function_composition}
\begin{figure}[ht!]
	\captionsetup[subfigure]{justification=centering}
	\centering
	\begin{subfigure}[t]{0.23\textwidth}
		\centering
		\includegraphics[width=0.93\textwidth]{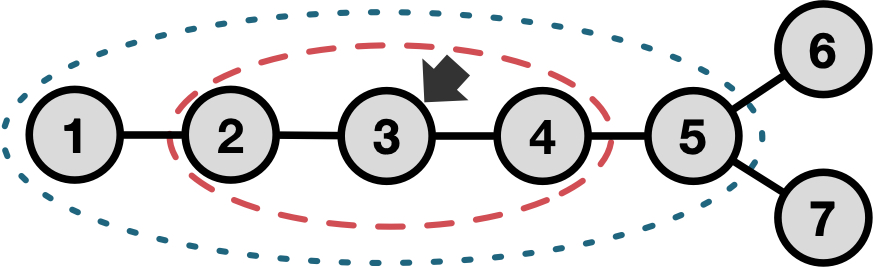}
		\caption{1- and 2-hop neighborhood of node 3}
		\label{fig_func_composition_a}
	\end{subfigure}
	~
	\begin{subfigure}[t]{0.26\textwidth}
		\centering
		\includegraphics[width=0.93\textwidth]{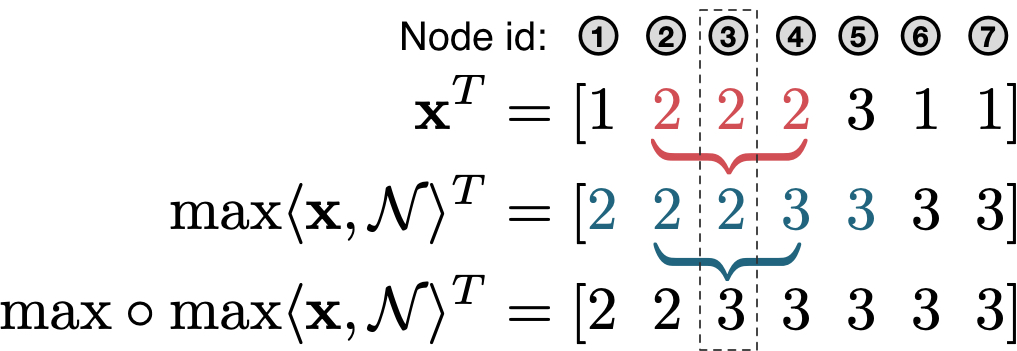}
		\caption{$\max\circ\max\langle\vx, \neighborhood\rangle$ captures the max degree in 2-hop neighborhood}
		\label{fig_func_composition_b}
	\end{subfigure}
	\vspace{-0.3cm}
	\caption{The composition of relational functions incorporates node degrees (column vector $\mathbf{x}$) in expanded subgraphs. 
	}
	\label{fig_func_composition}
	\vspace{-.45cm}
\end{figure}

To derive complex \& non-linear node features automatically, \method iteratively applies operators $\phi\in\Phi$ (e.g., \textit{mean}, \textit{variance}, \textit{sum}, \textit{max}, \textit{min}, \textit{l2-distance}) to lower-order functions, resulting in function compositions. $\ell$ such compositions of functions over a node's egonet $\neighborhood_i$ captures \textit{higher-order structural features} associated with the $\ell-$hop neighborhoods.
For example, assuming $\mathbf{x}$ is the vector consisting of node-wise degrees, the \textit{max} operator captures the maximum degree in the neighborhood $\neighborhood$ of a node. The application of the \textit{max} operator to all the nodes forms a new feature vector $\max\langle\vx, \neighborhood\rangle$ where each entry records the maximum degree in the corresponding neighborhood. Fig.~\ref{fig_func_composition} shows that the maximum degree of node $\{2,3,4\}$ is aggregated for node 3 in $\max\langle\vx, \neighborhood\rangle$ By iteratively applying \textit{max} to $\max\langle\vx, \neighborhood\rangle$ in the same way, the maximum value from broader neighborhood $\neighborhood$ is aggregated, which is equivalent to finding the maximum degree in the $2$-hop neighborhood. Fig.~\ref{fig_func_composition_b} depicts this process for node 3.

Formally, at level $\ell\in\{1, \dots, L\}$, a new function is composed as:
\begin{equation}
f^{(\ell)} = \phi\circ f^{(\ell-1)}, \forall{\phi\in\Phi}
\label{eq_func_comp}
\end{equation}
where $L < \texttt{diam}(G)$ or the diameter of $G$, and $f^{(0)}=f_b$(\S~\ref{sec:base-functions}).
We formally define some operators $\phi\in\Phi$ in Appendix~\ref{sec_exp_conf}.
Applying $f^{(\ell)}$ to $\mX^{(0)}$ generates order-$\ell$ structural features of the graph as $\mX^{(\ell)}$. 
In practice, \method recursively generates $\mX^{(\ell)}$ from $\mX^{(\ell-1)}$ by applying a total of $|\Phi|$ operators.  
The particular order in which relational operators are applied records how a function is generated.
\method then collects the composed relational functions per level into $\Fr$ as a part of the latent summary.

In terms of space, Equation~\eqref{eq_func_comp} indicates the dimension of $\mathcal{F}_r$ grows exponentially with $|\Phi|$, \ie, $|\Fr^{(\ell)}|=|\B||\Phi|^{(\ell)}$, which is also the number of columns in $\mathbf{F}^{(\ell)}$.
However, the max level $L$ is bounded with the diameter of $G$, that is $L \le \texttt{diam}(G)-1$ because functions with orders higher than that will capture the same repeated structural information. Therefore, the size of $\Fr$ is also bounded with $L$.
Although the number of relational functions grows exponentially, real-world graphs are extremely dense with small diameters $\texttt{diam}(G) \propto \log\log N$~\cite{cohen2003scale}.
In our experiments in \S~\ref{sec_exp}, $|\Fr|\approx 1000$ for $|\mathcal{B}|=3$ base functions, $|\Phi|=7$ operators, and $L=2$ levels.

\vspace{-0.1cm}
\subsection{Heterogeneous Context }
\label{sec_summarizing_hetero}

So far we have discussed how to obtain the base structural feature matrix $\mXo$ and the multi-level structural feature representations $\mXl$ by recursively employing the relational functions. As we show empirically in supplementary material~\ref{sec_hetero_context_justification}, directly deriving the structural embeddings based on these representations leads to low performance due to skewness in the extracted structural features. Here we discuss an intermediate transformation of the generated matrices that helps capture rich contextual patterns in the neighborhoods of each node, and eventually leads to a powerful summary.

\subsubsection{Handling skewness} 
\begin{figure}[t!]
	\centering
	\includegraphics[width=0.95\linewidth]{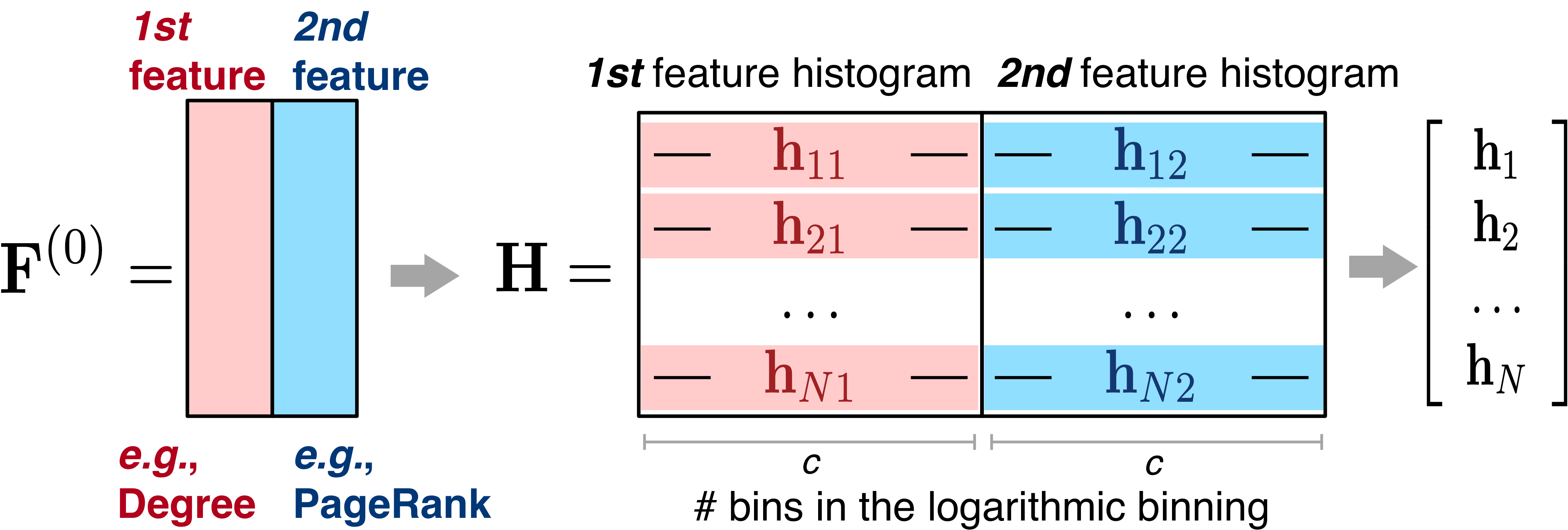}
	\vspace{-0.25cm}
	\caption{Example of creating histogram-based matrix representation $\mY^{(0)}$ with $Z=2$ features in the base feature matrix $\mX^{(0)}$. A single object / edge type and no edge directionality is assumed here for simplicity.}
	\label{fig:histogram}
	\vspace{-0.5 cm}
\end{figure}

For simplicity, we first discuss the case of a homogeneous network $G$ with a single node and edge type, and undirected edges. To handle the skewness in the higher-order structural features (\S~\ref{sec_multi_level_structure_extraction}) and more effectively capture the structural identity of each node within its context (i.e., non-typed neighborhood), we opt for an intuitive approach: for each node $i$ and each base/higher-order feature $j$, we create a histogram $\hist{ij}$ with $c$ bins for the nodes in its neighborhood $\neighborhood_i$. Variants of this approach are used to capture node context in existing representation learning methods, such as struc2vec~\cite{struc2vec} and xNetMF~\cite{regal}. 
In our setting, the structural identity of node $i$ is given as the concatenation of all its feature-specific histograms.
\begin{equation}
\hist{i} = \big[\,\hist{i1} \;\, \hist{i2} \; \cdots \; \hist{iZ}\,\big],    
\label{eq:identity}
\end{equation}
where $Z=\Bsize + \sum_{\ell=1}^{L} \Bsize \cdot \Phisize^{\ell}$ is the total number of histograms, or the number of base and higher-order features.  
Each histogram is in logarithmic scale to better describe the power-law-like distribution in graph features and has a total of $c$ bins.
By stacking all the nodes' structural identities vertically, we obtain a rich histogram-based context matrix $\mY = \big[\,\hist{1};\, \hist{2}; \cdots ; \hist{N} \big]$ as shown in Fig.~\ref{fig:histogram}.

\subsubsection{Handling object/edge types and directionality} The histogram-based representation that we described above can be readily extended to handle any arbitrary network $G$  with multiple object types, edge types and directed edges {\bf(P1)}. The idea is to capture the structural identity of each node $i$ within its \textit{different} contexts: 
\begin{itemize}
	\item $\neighborhood_i^t$ or $\neighborhood_i^{\tau}$: the neighborhood that contains only nodes of type $t\in\T_V$ or edges of type $\tau\in\T_E$, and
	\item $\neighborhood_i^+$ or $\neighborhood_i^-$: the neighborhood with only outgoing or incoming edges for node $i$. \end{itemize}
For example, to handle different object types, we create a context matrix $\mY^t_o$ by restricting the histograms on neighbors of type $t$, $\neighborhood_i^t$. These per-type matrices can be stacked into a tensor $\underline{\mathcal{H}}$, with each slice corresponding to a node-level histogram, $\mY^{t}_{o}$ of object type, $t$. Alternatively, the tensor can be matricized by frontal slices. By further restricting the neighborhoods to contain specific edge types and/or directionality in a similar manner, we can obtain the histogram-based representations $\mY^t_e$ and $\mY^t_d$, respectively. 

By imposing all of the restrictions at once, we can also obtain context matrix $\mY$ that accounts for all types of heterogeneity. 
We discuss this step with more details that may be necessary for reproducibility in \S~\ref{app_summarizing_hetero} of the supplementary material.

\vspace{-0.1cm}
\subsection{Latent Summarization} \label{sec_summarization}
\providecommand{\DLoss}[2]{\ensuremath{\mathbb{D}_{\ell}\big(\,#1\, \| #2\big)}}

The previous two subsections can be seen as a general framework for automatically extracting, linear and non-linear, higher-order structural features that constitute the nodes' contexts at multiple levels $\ell$.
Unlike embedding methods that generate graph-size dependent node representations,
we seek to derive a compressed latent representation of $G$ {\bf(P2)} that supports on-the-fly generation of node embeddings and (inductive) downstream tasks {\bf(P3,P4)}.
Although graph summarization methods~\cite{liu2018graph} are relevant as they represent an input graph with a summary or supergraph, it is infeasible to generate latent node representations due to the incurred information loss. Thus, such methods, which have different end goals, do not satisfy {\bf(P4)}.

\subsubsection{Multi-level Summarization}
\method explores node similarity based on the assumption that similar nodes should have similar structural context over neighborhoods of different hops.
Given the histogram-based context matrix $\mY^{(\ell)}$ that captures the heterogeneity of feature values associated with the $\ell-$order egonets in $G$ (\S~\ref{sec_function_composition}),
\method obtains the level-$\ell$ summarized representation $\mH^{(\ell)}$ via factorization  
$\mY^{(\ell)}=\mathbf{Y}^{(\ell)}\mH^{(\ell)}$, where $\mathbf{Y}^{(\ell)}$ is the dense node embedding matrix that we do \textit{not} store. Then, the latent summary $\mathcal{J}$ consists of the set of relational functions $\Fr$ (\S~\ref{sec_multi_level_structure_extraction}), and the multi-level summarized representations $\mathcal{S}=\{\mH^{(1)}, \ldots, \mH^{(\ell)}\}$. Though any technique can be used (e.g., NMF), we give the factors based on SVD for illustration: 
\vspace{-0.3cm}
\begin{align}
	& \text{\small \tt level-$\ell$ node embeddings (\underline{not stored}):}\;  \mathbf{Y}^{(\ell)} = \mathbf{U}^{(\ell)}\sqrt{\Sigma^{(\ell)}}\\
	& \text{\small \tt level-$\ell$ summarized representation:}\;\;\;\;\; \mH^{(\ell)} = \sqrt{\Sigma^{(\ell)}}\mV^{(\ell)T}
	\label{eq_summarize}
	\vspace{-0.5cm}
\end{align}
where $\Sigma^{(\ell)}$ are the singular values of $\mY^{(\ell)}$, and $\mU^{(\ell)T}$, $\mV^{(\ell)T}$ are its left and right singular vectors, respectively.

Intuitively, $\mH^{(\ell)}$ contains the best-fit $K^{(\ell)}$-dimensional subspace vectors for node context $\mY^{(\ell)}$ in the neighborhood at order-$\ell$.
The summary representations across different orders form the hierarchical summarization of $G$ that contains both local and global structural information, and the derived embedding matrix $\mathbf{Y}^{(\ell)}$ also preserves node similarity at multiple levels.
There is no need to store any of the intermediate matrices $\mX^{(\ell)}$ and $\mY^{(\ell)}$, nor the node embeddings $\mathbf{Y}^{(\ell)}$. The former two matrices can be derived on the fly given the composed relational functions $\Fr$.
Then, the latter can be efficiently estimated using the obtained sparse $\mY^{(\ell)}$ matrix and the stored summarized matrix $\mH^{(\ell)}$ through SVD (\S~\ref{sec_theoretical-analysis} gives more details).
Moreover, since the elements of the summary $\mathcal{S}$, i.e., the relational functions $\Fr$ and the factorized matrices, are independent of the nodes or edges of the input graph, both require trivial storage and achieve compression efficiency {\bf (P2)}.
We provide the pseudo-code of \method in Algorithm~\ref{alg:rel-func-learning-framework}.

We note that the relational functions $\mathcal{F}_r$ are a key enabling factor of our summarization approach.  
Without them, other embedding methods cannot benefit from our proposed summarized representations $\mathcal{S}$, nor reconstruct the node context and embeddings.

\subsubsection{Inductive Summaries {\bf(P3)}} \label{sec:framework-inductive}
The higher-order features derived from the set of relational functions $\Fr$ are structural, and thus generalize across graphs~\cite{rolX,regal,role2vec} and are independent of node IDs. As such, the factorized matrices in $\mathcal{S}$ learned on $G$ can be transferred to another graph $G^{\prime}$ to learn the node embeddings $\mathbf{Y}^{(\ell)}$ of a new, previously unseen graph $G^{\prime}$ as:

\vspace{-0.2cm}
\begin{equation}
{\mathbf{Y}^{\prime}}^{(\ell)} = {\mY^{\prime}}^{(\ell)}({\mH^{(\ell)}})^{\dagger}
\label{eq_inductive}
\vspace{-0.1cm}
\end{equation}
where $\mH^{(\ell)}\in\mathcal{S}$ is learned on $G$, ${\dagger}$ denotes the pseudo-inverse, and ${\mY^{\prime}}^{(\ell)}$ is obtained via applying $\Fr$ to $G^{\prime}$. The pseudo-inverse, $({\mH^{(\ell)}})^{\dagger}$ can be computed efficiently through SVD as long as the rank of $\mH^{(\ell)}$ is limited (\eg, empirically setting $K^{(\ell)}\le 128$)~\cite{brand2006fast}.

Equation~\eqref{eq_inductive} requires the same dimensionality $K^{(\ell)}=K^{\prime(\ell)}$ and the same number of bins of histogram context matrices $c=c^{\prime}$ at each  level $\ell$. The embeddings learned inductively  reflect the node-wise structural difference between graphs, $G$ and $G^{\prime}$, which can be used in applications of graph mining and time-evolving analysis. We present an application of temporal event detection in \S~\ref{sec:exp-anomaly-detection}. 

\setlength{\textfloatsep}{6pt}
\algrenewcommand{\alglinenumber}[1]{\fontsize{7}{9}\selectfont#1:}
\algblockdefx[parallel]{parfor}{endpar}[1][]{$\textbf{parallel for}$ #1 $\textbf{do}$}{$\textbf{end parallel}$}
\algtext*{endpar}\algrenewcommand{\alglinenumber}[1]{\fontsize{6.5}{11}\selectfont#1:}
\begin{algorithm}[t!]
	\caption{\,\small \fontsize{9}{10.5}\selectfont
		\method
	}
	\label{alg:rel-func-learning-framework}
	{\begin{spacing}{1.15}
			\fontsize{7}{8}\selectfont
			\begin{algorithmic}[1]
				\fontsize{8}{9}\selectfont
				\Statex \textbf{Input:}
				(un)directed heterogeneous graph $G$, a set of relational operators $\mPhi$;
				layer-wise embedding dimensionality $K^{(\ell)}$, for $K=\sum^{L}_{\ell=1}K^{(\ell)}$ dimensions in total;
				number of bins $c$ for the histogram representation \Statex \textbf{Output:} Summary $\mathcal{J}=\big\{\mathcal{F}, \mathcal{S}\big\}$
				\medskip
				\fontsize{8}{9}\selectfont

				\State $\Fr \leftarrow f_b$ \Comment{\texttt{\textcolor{gray}{Base graph functions: Eq.~\eqref{eq_base_vector}}}}
				\State Initialize $\mX^{(0)}$\Comment{\texttt{\textcolor{gray}{Base feature matrix Eq.~\eqref{eq_x0}}}}
				\label{algline:typed-base-functions}
				
				\For{$\ell=1,\!\ldots,\!L$} \label{algline:for-layer}
				\Comment{\texttt{\textcolor{gray}{multi-level summarization}}}
				\vspace{0.6mm}

				\For{$i=1,\!\ldots,\!|\Phi|$} \label{algline:for-rel-op}
				\label{algline:rel-operator}
				\Comment{\texttt{\textcolor{gray}{relational operators}}}                \vspace{0.6mm}
				
				\parfor[$j=1,\!\ldots,\!BR^{\ell-1}$]
				\Comment{\texttt{\textcolor{gray}{Columns in $\mX^{(l)}$}}}
				\label{algline:for-each-col-j}
				\vspace{0.8mm}
				
				\State $f = \phi_{i}\circ f_{j}^{(\ell-1)}$
				\Comment{\texttt{\textcolor{gray}{Compose func. in $\Fr^{(\ell-1)}$}}}
				\State $\mX^{(\ell)} = \mX^{(\ell)} \cup \phi_i \langle \mX_{:,j}^{(\ell-1)}, \neighborhood \rangle$
				\label{algline:compute-feature-vector-node-i} 
				\Comment{\texttt{\textcolor{gray}{Feature concatenation}}}
				\label{algline:compute-feature-vector-col-j} 
				\EndPar

				\EndFor
				\State Derive heterogeneous context $\mY^{(\ell)}$ \Comment{\texttt{\textcolor{gray}{\S~\ref{sec_summarizing_hetero} and Eq.~\eqref{eq_binning_et}}}}
				
				\State {$\mH^{(\ell)} = \sqrt{\Sigma^{(\ell)}}\mV^{(\ell)T}$}
				\Comment{\texttt{\textcolor{gray}{SVD: $\mY^{(\ell)} = \mU^{(\ell)}\Sigma^{(\ell)}\mV^{(\ell)T}$}}}
				\label{algline:svd}
				\vspace{0.8mm}
				\State $\Fr \leftarrow \Fr \cup f$, $\mathcal{S} \leftarrow \mathcal{S} \cup \mH^{(\ell)}$

				\EndFor

				\label{alg-rel-func-learning-framework}
			\end{algorithmic}
		\end{spacing}
		\vspace{1.12mm}
	}
\end{algorithm}

\subsubsection{On-the-fly embedding derivation {\bf(P4)}} \label{sec:framework-on-the-fly}
Given the summarized matrix $\mH^{(\ell)}$ at level $\ell$, the embeddings of specific nodes that are previously seen or unseen can be derived efficiently.
\method first applies $\Fr$ to derive their heterogeneous context $\mY^{(l)}_{\text{sub}}$ based on the graph structure, and then obtains the embeddings via Eq.~\eqref{eq_inductive}.
We concatenate $\mathbf{Y}^{(\ell)}$ given as output at each level to form the final node embeddings~\cite{line}.
Given that the dimension of embeddings is $K^{(\ell)}$ at level $\ell$, the final embedding dimension is  $K=\sum^{L}_{\ell=1}K^{(\ell)}$.

\subsection{Generalization}

\label{sec:generalization}
Here we discuss the generalizations of our proposed approach to labeled and attributed graphs.
It is straightforward to see that homogeneous, bipartite, signed, and labeled graphs are all special cases of heterogeneous graphs with 
$|\mathcal{T}_V|=|\mathcal{T}_E|=1$ types, 
$|\mathcal{T}_V|=2$ and $|\mathcal{T}_E|=1$ types, 
$|\mathcal{T}_V|=1$ and $|\mathcal{T}_E|=2$ types, and
$|\mathcal{T}_V|=\text{\#(node labels)}$ and $|\mathcal{T}_E|=\text{\#(edge labels)}$ types, respectively.
Therefore, our approach naturally generalizes to all of these graphs.
Other special cases include k-partite and attributed graphs.

\method also supports attributed graphs that have multiple attributes per node or edge (instead of a single label): Given an initial set of attributes organized in an attribute matrix $\mX_{a}$, we can concatenate $\mX_{a}$ with the base attribute matrix and apply our approach as before.  
Alternatively, we can transform the graph into a labeled one by applying a labeling function $\chi : \vx \rightarrow y$ that maps every node's attribute vector $\vx$ to a label $y$~\cite{role2vec}. Besides, our proposed method is easy to parallelize as the relational functions are applied to the subgraphs of each node independently, and the feature values are computed independently.

\subsection{Complexity Analysis}
\label{sec_theoretical-analysis}

\subsubsection{Computational Complexity.}\label{sec_computational_complexity}
\method is linear to the number of nodes $N$ and edges $M$. Per level, it derives the histogram-based context matrix $\mY^{(\ell)}$ and performs a rank-$K^{(\ell)}$ approximation.

\begin{Lemma}
	\label{lemma_computational_complexity}
	The computational complexity of \method is \\
	\vspace{-0.35cm}
	\begin{center}$\mathcal{O}( (c|\Fr||\mathcal{T}_V| |\mathcal{T}_E| + K^2)N + M)$.\end{center} \vspace{-0.2cm}
\end{Lemma}
We give the proof in Appendix~\ref{sec_computation_complexity}. As indicated in \S~\ref{sec_multi_level_structure_extraction}, the number of features in $\mY^{(\ell)}$ across $L$ layers is equivalent to the number of composed relational functions $|\Fr|$. Since $|\Fr|$ is bounded with $L$ and $L<\texttt{diam}(G)$, the term $(c|\Fr||\mathcal{T}_V| |\mathcal{T}_E| + K^2)$ forms a constant related to graph heterogeneity and structure.

\subsubsection{Space Complexity.}
The runtime and output compression space complexity of \method is given in Lemma~\ref{lemma_space_complexity}. In the runtime at level $\ell$, \method leverages $\mX^{(\ell-1)}$ to derive $\mX^{(\ell)}$ and $\mY^{(\ell)}$, which comprise two terms in the runtime space complexity.
We detail the proof in Appendix~\ref{sec_space_complexity}

\begin{Lemma}\label{lemma_space_complexity}
	The  \method space complexity during runtime is $\mathcal{O}( (c|\Fr||\mathcal{T}_V| |\mathcal{T}_E| + |\Fr|)N)$.
	The space needed for the {\em output} of \method is $\mathcal{O}(cK|\Fr||\mathcal{T}_V||\mathcal{T}_E| + |\Fr|).$
\end{Lemma}
The output of \method that needs to be stored (i.e., set of relational functions $\Fr$ and summary matrices in $\mathcal{S}$) is independent of $N, M$. 
Compared with output embeddings with complexity $\mathcal{O}(NK)$ given by existing methods, \method satisfies the crucial property we desire {\bf(P2)} from latent summarization (Def.~\ref{def_latent_graph_summarization}).

\section{Experiments} 
\label{sec_exp}

In our evaluation we aim to answer four research questions: \begin{itemize}
	\item[\bf Q1] How much space do the \method summaries save {(P2)}?
	\item[\bf Q2] How does \method     perform in machine learning tasks, such as link prediction in heterogeneous graphs {(P1)}?
	\item[\bf Q3] How well does it perform in inductive tasks {(P3)}?
	\item[\bf Q4] Does \method scale well with the network size? 
\end{itemize}
We have discussed on-the-fly embedding derivation (P4) in \S~\ref{sec:framework-on-the-fly}.

\subsection{Experimental Setup} \label{sec:exp-data}

\subsubsection{Data} In accordance with {\bf (P1)}, we use a variety of real-world heterogeneous network data from Network Repository~\cite{nr}.
We present their statistics in Table~\ref{table_stats_t}.
\begin{itemize}    
	\item \textbf{Facebook}~\cite{node2vec} is a homogeneous network that represents friendship relation between users. 
	
	\item \textbf{Yahoo! Messenger Logs}~\cite{deepGL} is a heterogeneous network of Yahoo! messenger communication patterns, where edges indicate message exchanges. The users are associated with the locations from which they have sent messages. 
	
	\item \textbf{DBpedia}\footnote{\url{http://networkrepository.com/}
		\label{dataset_footnote}} is an unweighted, heterogeneous subgraph from DBpedia project consisting of 4 types of entities and 3 types of relations: user-occupation, user-work ID, work ID-genre.     
	
	\item \textbf{Digg}\footref{dataset_footnote} is a heterogeneous network that records the voting behaviors of users to stories they like. Node types include users and stories. Each edge represents one vote or a friendship.
	
	\item \textbf{Bibsonomy}\footref{dataset_footnote} is a k-partite network that represents the behaviors of users assigning tags to publications. 
\end{itemize}

\subsubsection{Baselines} We compare \method with baselines commonly used in graph summarization, matrix factorization and representation learning over networks, namely, they are:
{\bf (1)}~Node aggregation or NA for short~\cite{zhu2016unsupervised,blondel2008fast},
{\bf (2)}~Spectral embedding or SE~\cite{spectral},
{\bf (3)}~LINE~\cite{line},
{\bf (4)}~DeepWalk or DW~\cite{perozzi2014deepwalk}, 
{\bf (5)}~Node2vec or n2vec~\cite{node2vec},  
{\bf (6)}~struc2vec or s2vec~\cite{struc2vec}, 
{\bf (7)}~DNGR~\cite{cao2016deep}, 
{\bf (8)}~GraRep or GR~\cite{cao2015grarep}, 
{\bf (9)}~Metapath2vec or m2vec~\cite{dong2017metapath2vec}, and
{\bf (10)}~AspEm~\cite{shi2018aspem},
{\bf (11)}~Graph2Gauss or G2G~\cite{bojchevski2018deep}.
To run baselines that do not explicitly support heterogeneous graphs, we align nodes of the input graph according to their object types and re-order the IDs to form the homogeneous representation. 
In node aggregation, CoSum~\cite{zhu2016unsupervised} ran out of memory due to the computation of pairwise node similarity. We use Louvain~\cite{blondel2008fast} as an alternative that scales to large graphs and forms the basis of many node aggregation methods.

\subsubsection{Configuration}
We evaluate \method with $L=1$ and $L=2$ to capture subgraph structural features in 1-hop and 2-hop neighborhoods, respectively, against the optimal performance achieved by the baselines.
We derive in-/out- and total degrees to construct the $N\times 3$ base feature matrix $\mX^{(0)}$.
Totally, we generate $\approx 1000$ composed functions, each of which corresponds to a column vector in $\mX$.
For fairness, we do not employ parallelization and terminate processes exceeding 1 day. The output dimensions of all node representations are set to be $K=128$. 
We also provide an ablation study in terms of the choice of initial vectors, different sets of relational operators in  supplementary material~\ref{sec_hetero_context_justification}-\ref{sec_justification_operators}. For reproducibility, we detail the configuration of all baselines and \method in Appendix~\ref{sec_exp_conf}. The source code is available at \url{https://github.com/GemsLab/MultiLENS}.

\begin{table}[t!]
	\centering
	\caption{Statistics for the heterogeneous networks that we use in our experiments.
	}
	\vspace{-0.1cm}
	\centering 
	{\small
		\setlength{\tabcolsep}{6pt} \def\arraystretch{1.1} \begin{tabularx}{\linewidth}{@{}lrrcr@{}}
			\toprule
			\textbf{Data} & \#Nodes & \#Edges & \#Node Types & Graph Type
			\\ \hline
			facebook & 4\,039 &  88\,234 & 1 & unweighted \\ yahoo-msg & 100\,058 & 1\,057\,050 & 2 & weighted \\ dbpedia & 495\,936 & 921\,710 & 4 &  unweighted \\ digg & 283\,183 & 4\,742\,055 & 2 & unweighted\\ bibsonomy & 977\,914  & 3\,754\,828 & 3 & weighted\\
			\bottomrule
		\end{tabularx}
	}
	\label{table_stats_t}
\end{table}

\subsection{ Compression rate of \method}
\label{sec:exp-graph-summarization}
The most important question for our latent summarization method (Q1) is about how well it compresses large scale heterogeneous data {\bf (P2)}. To show \method's benefits over existing embedding methods, we measure the storage space for the generated embeddings by the baselines that ran successfully. In Table~\ref{table:exp-graph-summarization-vs-embedding} we report the space required by the \method summaries in MB, and the space that the outputs of our baselines require \textit{relative} to the corresponding \method summary. We observe that the latent summaries generated by \method take up very little space, well under 1MB each. The embeddings of the representation learning baselines take up $80-2152\times$ more space than the \method summaries on the larger datasets. On Facebook, which is a small dataset with 4K nodes, the summarization benefit is limited; the baseline methods need about $3-12\times$ more space. In addition, the node-aggregation approach takes up to $12\times$ storage space compared to our latent summaries, since it generates an $N\times 1$ vector that depends on graph size to map each node to a supernode.  
This further demonstrates the advantage of our graph-size independent latent summarization.

\begin{table}[bh!]
	\centering
	\vspace{-0.4cm}
	\caption{Output storage space required for embedding methods relative to the \method summaries (given in MB). 
		\method requires $3-2152\times$ less output storage space than embedding methods. }
	\vspace{-2mm}
	\centering 
	\fontsize{7}{7.5}\selectfont
	\setlength{\tabcolsep}{1.5pt} \label{table:exp-graph-summarization-vs-embedding}
	\vspace{-0.1cm}
	\def\arraystretch{1.3} \begin{tabular}{l H C{0.75cm} C{0.8cm} C{0.8cm} C{0.75cm} C{0.7cm} C{0.8cm} C{0.75cm} C{0.8cm}}
		\toprule
		\textbf{Data} & NA & SE & LINE & n2vec & DW & m2vec & AspEm & G2G & {\bf ML} (MB)
		\\ \midrule
		facebook & - & 8.13x & 8.48x & 12.79x & 12.84x & 3.82x & 8.50x & 9.17x & \textbf{0.58} \\
		yahoo & 1.4x & 187.1x & 180.0x & 242.2x & 231.0x & 79.8x & 197.4x & 195.8x & \textbf{0.62} \\
		dbpedia & 6.3x & 710.0x & 714.2x & 996.4x & 996.2x &  - & 749.2x & 743.6x & \textbf{0.81} \\
		digg & 4.7x & 608.2x & 612.8x & 848.9x & 830.3x & 259.9x & 641.7x & 635.2x & \textbf{0.54} \\
		bibson. & 12.3x & 1512.1x & 1523.0x & 2152.5x & 2152.5x & - & 1595.8x & - & \textbf{0.75} \\
		
		\bottomrule
	\end{tabular}
	\vspace{-0.3cm}
\end{table}

\subsection{Link Prediction in Heterogeneous Graphs}
\label{sec:exp-link-pred}

\begin{table*}[h!]
	{\footnotesize
		\centering
		\caption{Link prediction: node embeddings derived by \method (ML) outperforms all baselines measured by every evaluation metric. Specifically, \method outperforms embedding baselines by $3.46\%\sim34.34\%$ in AUC and $3.71\%\sim31.33\%$ in F1 on average. It outperforms even more over the aggregation-based methods. 
			The asterisk $^{\ast}$ denotes statistically significant improvement over the best baseline at $p<0.01$ in a two-sided t-test.
			OOT = Out Of Time (12 hours),  OOM = Out Of Memory (16GB).
		}
		\label{table_link_prediction}
		\vspace{-2mm}
		\def\arraystretch{0.8}
		\begin{tabular}{lr C{0.7cm} C{0.7cm} C{0.8cm} C{0.7cm} C{0.8cm} C{0.7cm} C{0.8cm} C{0.7cm} C{0.7cm} C{0.75cm} H C{0.7cm} | C{1.1cm} C{1.15cm}}
			\toprule
			Data & Metric & NA & SE & LINE & DW & n2vec & GR & s2vec  &
			DNGR & m2vec & AspEm & GCN & G2G &
			{\bf ML}($L=1$) & {\bf ML}($L=2$)
			\\ \midrule
			facebook
			& \begin{tabular}{@{}r@{}}\textbf{AUC} \\ \textbf{ACC}  \\\textbf{F1 macro}\end{tabular}
			& \begin{tabular}{@{}c@{}} 0.6213 \\ 0.5545 \\ 0.5544 \end{tabular}
			& \begin{tabular}{@{}c@{}} 0.6717 \\ 0.5995  \\ 0.5716 \end{tabular}
			& \begin{tabular}{@{}c@{}} 0.7948 \\ 0.7210  \\ 0.7210 \end{tabular}
			& \begin{tabular}{@{}c@{}} 0.7396 \\ 0.6460 \\ 0.6296 \end{tabular}
			& \begin{tabular}{@{}c@{}} 0.7428 \\ 0.6544  \\ 0.6478 \end{tabular}
			& \begin{tabular}{@{}c@{}} 0.8157 \\ 0.7368  \\ 0.7367 \end{tabular}
			& \begin{tabular}{@{}c@{}} 0.8155 \\ 0.7388  \\ 0.7387 \end{tabular}
			& \begin{tabular}{@{}c@{}} 0.7894 \\ 0.7062  \\ 0.7060 \end{tabular}
			& \begin{tabular}{@{}c@{}} 0.7495 \\ 0.7051   \\ 0.7041 \end{tabular}
			& \begin{tabular}{@{}c@{}} 0.5886 \\ 0.5628  \\ 0.5628 \end{tabular}
			& OOM & \begin{tabular}{@{}c@{}} 0.7968 \\ 0.7274  \\ 0.7273 \end{tabular}
			& \begin{tabular}{@{}c@{}}0.8703 \\{\bf 0.7920}$^{\boldmath{\ast}}$ \\{\bf 0.7920}$^{\boldmath{\ast}}$ \end{tabular} 
			& \begin{tabular}{@{}c@{}}{\bf 0.8709$^{\boldmath{\ast}}$} \\ 0.7904  \\ 0.7905\end{tabular} 
			\\\midrule
			
			yahoo-msg
			& \begin{tabular}{@{}r@{}}\textbf{AUC} \\ \textbf{ACC} \\\textbf{F1 macro}\end{tabular}
			& \begin{tabular}{@{}c@{}} 0.7189 \\ 0.2811  \\ 0.2343 \end{tabular}
			& \begin{tabular}{@{}c@{}} 0.5375 \\ 0.5224 \\ 0.5221 \end{tabular}
			& \begin{tabular}{@{}c@{}} 0.6745 \\ 0.6269 \\0.6265\end{tabular}
			& \begin{tabular}{@{}c@{}} 0.7715 \\ 0.6927  \\0.6897 \end{tabular}
			& \begin{tabular}{@{}c@{}} 0.7830 \\ 0.7036  \\0.7016 \end{tabular}
			& \begin{tabular}{@{}c@{}} 0.7535 \\ 0.6825  \\ 0.6821 \end{tabular}
			& OOT
			& OOM
			& \begin{tabular}{@{}c@{}}0.6708 \\ 0.6164  \\0.6145 \end{tabular}
			& \begin{tabular}{@{}c@{}}0.5587 \\ 0.5379  \\0.5377 \end{tabular}
			& OOM
			& \begin{tabular}{@{}c@{}} 0.6988 \\ 0.6564  \\ 0.6562\end{tabular}
			& \begin{tabular}{@{}c@{}}0.8443\hspace{0.1cm} \\ {\bf 0.7587}$^{\boldmath{\ast}}$  \\ {\bf 0.7577}$^{\boldmath{\ast}}$\end{tabular} 
			& {\bf \begin{tabular}{@{}c@{}}0.8446$^{\boldmath{\ast}}$ \\ 0.7587$^{\boldmath{\ast}}$ \\ 0.7577$^{\boldmath{\ast}}$\end{tabular} }
			\\\midrule
			dbpedia
			& \begin{tabular}{@{}r@{}}\textbf{AUC} \\ \textbf{ACC}  \\\textbf{F1 macro}\end{tabular}
			& \begin{tabular}{@{}c@{}} 0.6002 \\ 0.3998  \\ 0.2968 \end{tabular}
			& \begin{tabular}{@{}c@{}} 0.5211 \\ 0.5399  \\ 0.4539 \end{tabular}
			& \begin{tabular}{@{}c@{}} 0.9632 \\ 0.9111  \\ 0.9110 \end{tabular}
			& \begin{tabular}{@{}c@{}} 0.8739 \\ 0.8436  \\ 0.8402 \end{tabular}
			& \begin{tabular}{@{}c@{}} 0.8774 \\ 0.8436  \\ 0.8402 \end{tabular}
			& OOM
			& OOT
			& OOM
			& OOT
			& \begin{tabular}{@{}c@{}} 0.6364 \\ 0.5869  \\0.5860  \end{tabular}
			& OOM
			& \begin{tabular}{@{}c@{}} 0.7384 \\ 0.6625  \\ 0.6613\end{tabular}
			& {\bf \begin{tabular}{@{}c@{}}0.9820$^{\boldmath{\ast}}$ \\  0.9186  \\ 0.9186\end{tabular} }
			& \begin{tabular}{@{}c@{}}0.9809 \\ 0.9151  \\ 0.9150\end{tabular} 
			\\\midrule
			digg
			& \begin{tabular}{@{}r@{}}\textbf{AUC} \\ \textbf{ACC}  \\\textbf{F1 macro}\end{tabular}
			& \begin{tabular}{@{}c@{}}0.7199 \\ 0.2801  \\  0.2660 \end{tabular}
			& \begin{tabular}{@{}c@{}} 0.6625 \\ 0.6512  \\ 0.6223 \end{tabular}
			& \begin{tabular}{@{}c@{}} 0.9405 \\ 0.8709  \\ 0.8709 \end{tabular}
			& \begin{tabular}{@{}c@{}} 0.9664 \\ 0.9023  \\ 0.9019 \end{tabular}
			& \begin{tabular}{@{}c@{}} 0.9681 \\ 0.9049  \\ 0.9046 \end{tabular}
			& OOM
			& OOT
			& OOM
			& \begin{tabular}{@{}c@{}} 0.9552 \\ 0.8891 \\ 0.8890 \end{tabular}
			& \begin{tabular}{@{}c@{}} 0.5644 \\ 0.5459  \\0.5459 \end{tabular}
			& OOM
			& \begin{tabular}{@{}c@{}} 0.8978 \\ 0.8492  \\ 0.8492\end{tabular}
			& {\bf \begin{tabular}{@{}c@{}}0.9894$^{\boldmath{\ast}}$ \\ 0.9596$^{\boldmath{\ast}}$  \\ 0.9595$^{\boldmath{\ast}}$\end{tabular} }
			& \begin{tabular}{@{}c@{}}0.9893 \\ 0.9590  \\ 0.9590\end{tabular}
			\\\midrule
			bibsonomy
			& \begin{tabular}{@{}r@{}}\textbf{AUC} \\ \textbf{ACC}  \\
				\textbf{F1 macro}\end{tabular}
			& \begin{tabular}{@{}c@{}} 0.7836 \\ 0.2164 \\ 0.2070 \end{tabular}
			& \begin{tabular}{@{}c@{}} 0.6694 \\ 0.6532 \\ 0.6064 \end{tabular}
			& \begin{tabular}{@{}c@{}} 0.9750 \\ 0.9350  \\ 0.9349 \end{tabular}
			& \begin{tabular}{@{}c@{}} 0.6172 \\ 0.5814  \\ 0.5781 \end{tabular}
			& \begin{tabular}{@{}c@{}} 0.6173 \\ 0.5816  \\ 0.5782 \end{tabular}
			& OOM
			& OOT
			& OOM
			& OOT
			& \begin{tabular}{@{}c@{}} 0.6127 \\ 0.5790 \\0.5772 \end{tabular}
			& OOM
			& OOM
			& {\bf \begin{tabular}{@{}c@{}} 0.9909$^{\boldmath{\ast}}$ \\  0.9485$^{\boldmath{\ast}}$ \\ 0.9485$^{\boldmath{\ast}}$ \end{tabular} }
			& \begin{tabular}{@{}c@{}} 0.9909 \\ 0.9466  \\ 0.9466\end{tabular} 
			\\
			\bottomrule
			
		\end{tabular}
		
		\vspace{-0.1cm}
	}
\end{table*}

For Q2, we investigate the performance of \method in link prediction task over heterogeneous graphs ({\bf P1}). 
We use logistic regression with regularization strength $= 1.0$ and stopping criteria$=10^{-4}$. 
An edge $e_{ij}$ is represented by the concatenating the embeddings of its source and destination: $emb(e_{ij}) = [emb(i), emb(j)]$ as used in~\cite{deepGL}.
For each dataset $G(V, E)$, we create the subgraph $G'(V, E')$ by keeping all the nodes but randomly removing $\sim40\%$ edges. 
We run all methods on $G'$ to get node embeddings and randomly select $10\%|E|$ edges as the training data.
Out of the removed edges, $25\%$ ($10\%|E|$) are used as missing links for testing. 
We also randomly create the same amount of ``fake edges'' for both training and testing. Table~\ref{table_link_prediction} illustrates the prediction performance measured with AUC, ACC, and F1 macro scores.

We observe that \method outperforms the baselines measured by every evaluation metric. \method outperforms embedding baselines by $3.46\%\sim34.34\%$ in AUC and $3.71\%\sim31.33\%$ in F1 score. 
For runnable baselines designed for node embeddings in homogeneous graphs (baseline {\bf 3 - 8}), the experimental result is expected as \method incorporates heterogeneous contexts within 2-neighborhood in the node representation. 
It is worth noting that \method outperforms Metapath2vec and AspEm, both of which are designed for heterogeneous graphs. One reason behind is the inappropriate meta-schema specified, as Metapath2vec and AspEm require predefined meta-path / aspect(s) in the embedding. 
On the contrary, \method does not require extra input and captures graph heterogeneity automatically.
We also observe the time and runtime space efficiency of \method when comparing with neural-network based methods (DNGR, G2G), GraRep and struc2vec on large graphs. Although the use of relational operators is similar to information propagation in neural-networks, \method requires less computational resource with promising results.
Moreover, the \method summaries for both $L=1$ and $L=2$ levels achieve promising results, but generally we observe that there is a slight drop in accuracy for higher levels. 
This indicates that node context at higher levels may incorporate noisy, less-relevant higher-order structural features (\S~\ref{sec_function_composition}).

\begin{figure*}[t!]
	\vspace{-0.2cm}
	\captionsetup[subfigure]{justification=centering}
	\centering
	\begin{subfigure}[t]{0.33\linewidth}
		\centering
		\includegraphics[width=\textwidth]{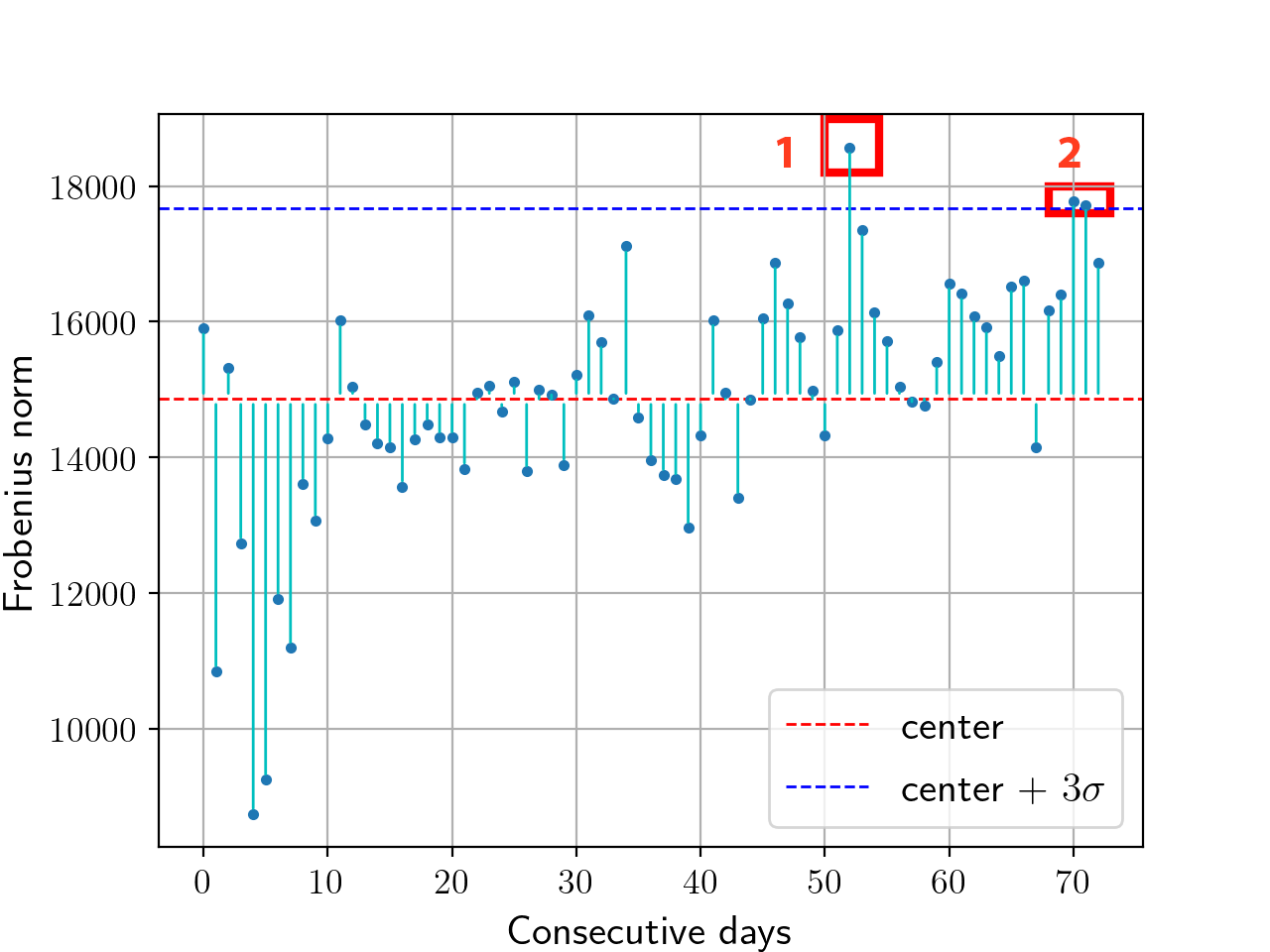}
		\caption{\texttt{Twitter}: Consecutive embeddings change in Twitter during 05/12/2014--07/31/2014. }
		\label{fig_real_a}
	\end{subfigure}
	~
	\begin{subfigure}[t]{0.33\linewidth}
		\centering
		\includegraphics[width=\textwidth]{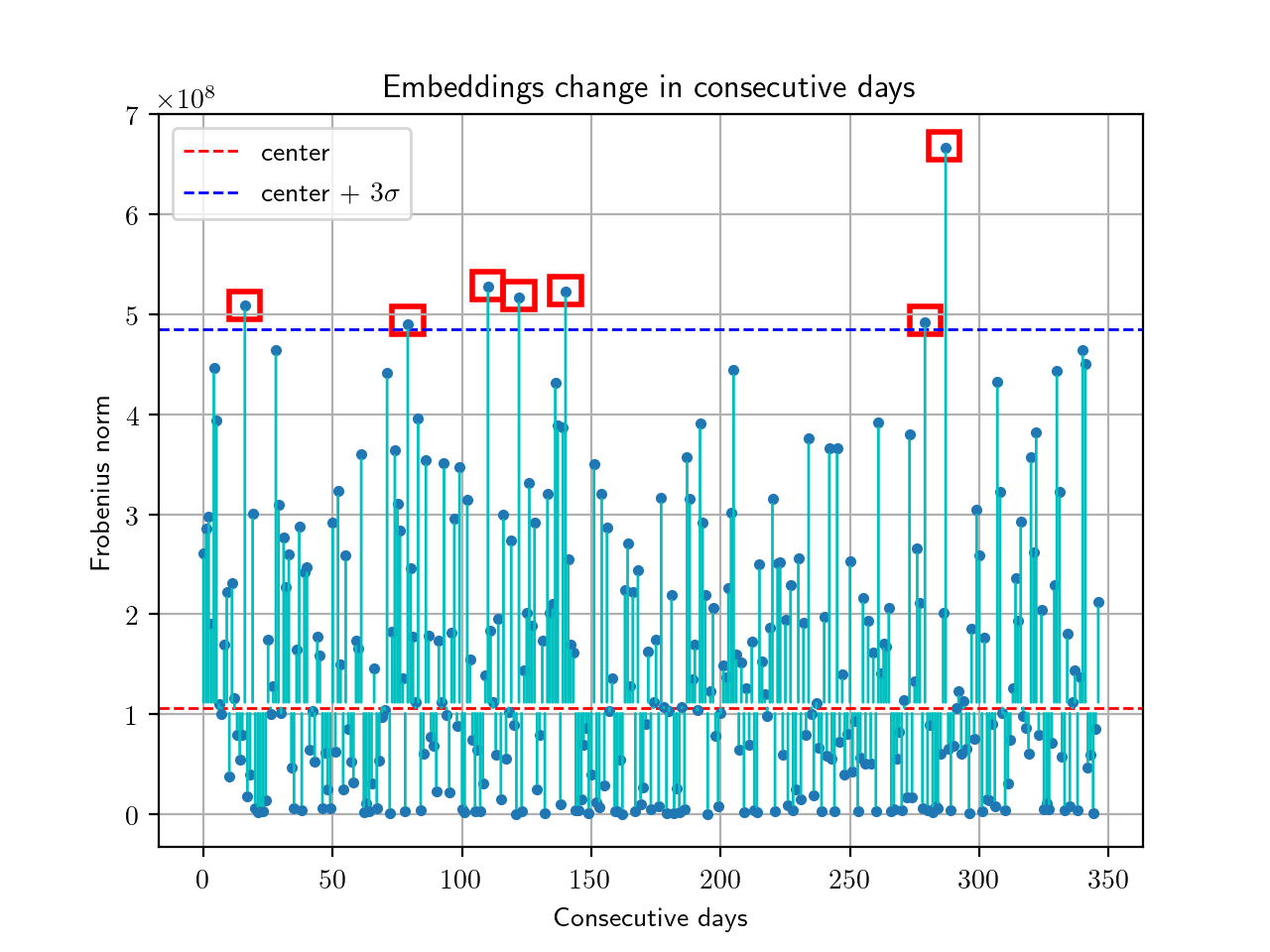}
		\caption{\texttt{Enron}: Consecutive embeddings change in weekdays during 01/01/2001--5/01/2002.}
		\label{fig_real_b}
	\end{subfigure}
	~
	\begin{subfigure}[t]{0.33\linewidth}
		\centering
		\includegraphics[width=\linewidth]{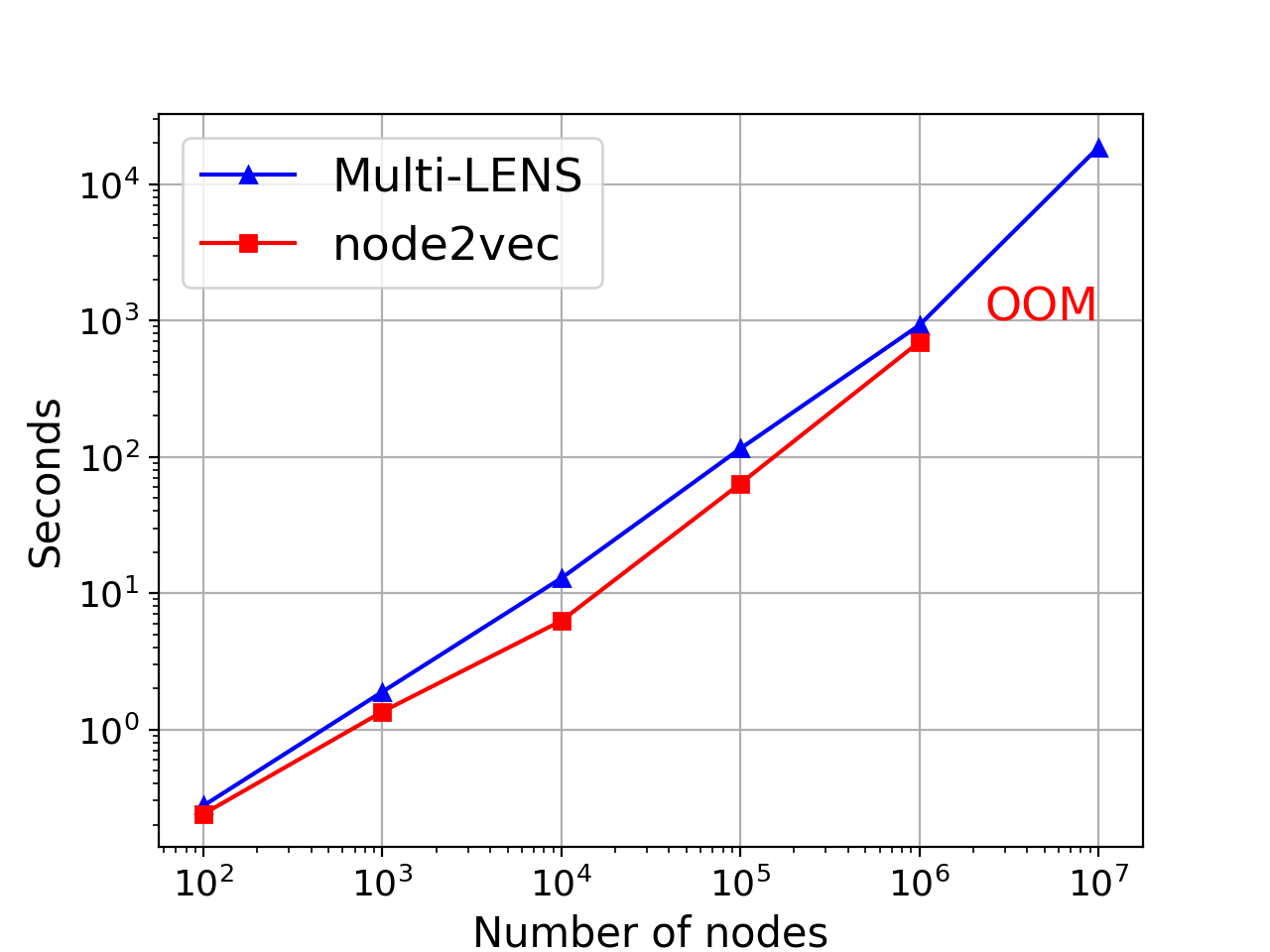}
		\caption{Runtime (in sec) for \method vs. node2vec.}
		\label{fig_scalability}
	\end{subfigure}
	\vspace{-0.3cm}
	\caption{(\subref{fig_real_a})-(\subref{fig_real_b}) Major event detection in real world datasets; (\subref{fig_scalability}) Runtime reported on ER graphs with $d_{\text{avg}}=10$. \method scales similarly to node2vec with less memory requirement while node2vec runs out of memory on the graph with $10^7$ nodes.}
	\label{AD_real}
	\vspace{-.3cm}
\end{figure*}

\begin{table}[b!]
	\centering
	{\small
		\caption{Anomalous Erd\H{o}s-R\'enyi (ER) subgraphs (with $n$ nodes and probability $p$) detection precision on both synthetic and real-world graphs. 
		}
		\vspace{-3mm}
		\label{table_anomaly_syn}
		\setlength{\tabcolsep}{5pt} \def\arraystretch{0.9} \begin{tabular}{@{}c cccccc !{\vrule width 0.6pt}  ccc }
			\noalign{\hrule height 0.7pt}
			\TT
			&& \multicolumn{5}{c !{\vrule width 0.6pt}}{\textbf{\sc Real Graph}} & \multicolumn{3}{c}{\textbf{\sc Synthetic Graph}}
			\\
			\BB
			& \textbf{p \textbackslash $\;$ n} & {\bf 100} & {\bf 200} & {\bf 300} & {\bf 400} & {\bf 500} & {\bf 50} & {\bf 75} & {\bf 100}\\
			\noalign{\hrule height 0.7pt}
			\TT
			& \textbf{0.1} & 0.200 & 0.780 & 0.950 & 0.973 & 0.980 & 0.06 & 0.3333 & 0.81\\
			& \textbf{0.3} & 0.870 & 0.960 & 0.990 & 0.995 & 0.996 & 1 & 1 & 1 \\
			& \textbf{0.5} & 0.920 & 0.990 & 0.993 & 1 & 1 & 1 & 1 & 1\\
			& \textbf{0.7} & 0.940 & 0.990 & 1 & 1 & 1 & 1 & 1 & 1\\
			& \textbf{0.9} & 0.980 & 1 & 1 & 1 & 1 & 1 & 1 & 1\\
			\noalign{\hrule height 0.7pt}
		\end{tabular}
	}
\end{table}

\vspace{-0.2cm}
\subsection{Inductive Anomaly Detection} \label{sec:exp-anomaly-detection} To answer Q3 about inductive learning, we first perform anomalous subgraph detection on both synthetic and real-world graphs. We also showcase the application of \method summaries on real-world event detection, in an inductive setting ({\bf P3}).
\vspace{-0.1cm}
\subsubsection{Anomalous Subgraph Detection}
Following the literature~\cite{miller2015spectral}, we first generate two ``background'' graphs, $G_{1}$ and $G_{2}$. We then induce an anomalous subgraph into $G_{2}$ by randomly selecting $n$ nodes and adding edges to form an anomalous ER subgraph with $p$ and $n$ shown in Table~\ref{table_anomaly_syn}.
We leverage the summary learned from $G_{1}$ to learn node embeddings in $G_{2}$, we identify the top-$n$ nodes with the highest change in euclidean distance as anomalies, and report the precision in Table~\ref{table_anomaly_syn}.  
In the synthetic setting, we generate two Erd\H{o}s-R\'enyi (ER) graphs, $G_1^{\text{syn}}$ and $G_2^{\text{syn}}$,  with $10^4$ nodes and average degree 10 ($p^{\text{back}}=10^{-3}$).
In the real-graph setting, we construct $G_{1}^{\text{real}}$ and $G_{2}^{\text{real}}$ using two consecutive daily graphs in the bibsonomy dataset.

In the synthetic scenario, we observe that \method gives promising results by successfully detecting nodes with the most deviating embedding values, except when the size of injection is small. In the case of very sparse ER injections ($p=0.1$), the anomalies are not detectable over the natural structural deviation between $G^{\text{syn}}_{1}$ and $G^{\text{syn}}_{2}$. However,  denser injections ($p\geq 0.3$) affect more significantly the background graph structure,
which in turn leads to notable change in the \method embeddings for the affected subset of nodes. For real-world graphs, we also observe that \method successfully detects anomalous patterns when the injection is relatively dense, even when the background graphs have complex structural patterns. This demonstrates that \method can effectively detect global changes in graph structures.

\vspace{-0.2cm}
\subsubsection{Graph-based Event Detection}
We further apply \method to real-world graphs to detect events that appear unusual or anomalous with respect to the global temporal behavior of the complex network.
The datasets we used are the \texttt{Twitter}\footnote{\url{http://odds.cs.stonybrook.edu/twittersecurity-dataset/}} and \texttt{Enron}\footnote{\url{http://odds.cs.stonybrook.edu/enroninc-dataset/}} graphs. \texttt{Twitter} has totally $308\,499$ nodes and $2\,601\,834$ edges lasting from 05/12/2014 to 07/31/2014, and \texttt{Enron} has totally $80848$ nodes and $2\,233\,042$ edges lasting from 01/01/2001 to 05/01/2002. Similar to the synthetic scenario, we split the temporal graph into consecutive daily subgraphs and adopt the summary learned from $G_{t-1}$ to get node embeddings of $G_{t}$. Intuitively, large distances between node embeddings of consecutive daily graphs indicate abrupt changes of graph structures, which may signal events.

Fig.~\ref{fig_real_a} shows the change of Frobenius norm between keyword / hashtag embeddings in consecutive instances of the daily \texttt{Twitter} co-mentioning activity. The two marked days are 3$\sigma$ (stdev) units away from the median value~\cite{koutra2013deltacon}, which correspond to serious events:
(1) the Gaza-Israel conflict and (2) Ebola Virus Outbreak. 
Compared with other events in the same time period, the detected ones are the most impactful in terms of the number of people affected, and the attraction they drew as they are related to terrorism or domestic security.
Similarly for \texttt{Enron}, we detect several events based on the change of employee embeddings in the Enron corpus from the daily message-exchange behavior. 
We highlight these events, which correspond to notable ones in the company's history, in Fig.~\ref{fig_real_b} and provide detailed information in Appendix~\ref{sec_exp_enron}.

\vspace{-0.2cm}
\subsection{Scalability of \method}
Finally, Q4 concerns the scalability of our approach. To that end, we generate Erd\H{o}s-R\'enyi graphs with average degree $d_\text{avg} = 10$, while varying the number of nodes from $10^2$ to $10^7$. For reference, we compare it against one of the fastest and most scalable baselines, node2vec. As shown in Fig.~\ref{fig_scalability}, node2vec runs out of memory on the graph with $10^7$ nodes, whereas \method scales almost as well as node2vec and to bigger graphs, while also using less space.

\section{Related Work} 
\label{sec_related-work}
We qualitatively compare \method to summarization and embedding methods in Table~\ref{table_comp}.

\subsubsection*{Node embeddings.} 
Node embedding or representation learning has been an active area which aims to preserve a notion of similarity over the graph in node representations~\cite{goyal2018graph,deepGL}. For instance, \cite{perozzi2014deepwalk, line, node2vec, cao2016deep, dong2017metapath2vec} define node similarity in terms of proximity (based on the adjacency or positive pointwise mutual information matrix) using random walks (RW) or deep neural networks (DNN). More relevant to \method are approaches capturing similar node behavioral patterns (roles) or structural similarity~\cite{roles2015-tkde,role2vec,regal,deepGL}. For instance, struc2vec and xNetMF~\cite{struc2vec,regal} define similarity based on node degrees, while DeepGL~\cite{deepGL} learns deep inductive relational functions applied to graph invariants such as degree and triangle counts.
\cite{levy2014neural, qiu2018network} investigate theoretical connection between matrix factorization and the skip-gram architecture.
To handle heterogeneous graphs, metapath2vec~\cite{dong2017metapath2vec} captures semantic and structural information by performing RW on predefined metapaths. 
There are also works based on specific characteristics in heterogeneous graphs. For example,
AspEm represents underlying semantic facets as multiple ``aspects'' and selects a subset to embed based on datasetwide statistics.
Unlike above methods that generate dense embeddings of fixed dimensionality, \method derives compact and multi-level latent summaries that can be used to generate node embeddings without specifying extra input.
The use of relational operators is also related to recent neural-network based methods.
For example, the \textit{mean} operator is related to the mean aggregator of GraphSAGE~\cite{hamilton2017inductive} and the propagation rule of GCN~\cite{kipf2017semi}. But unlike these and other neural-network based methods that propagate information and learn embeddings based on features from the local neighborhood,
\method learns latent subspace vectors of node contexts as the summary.

\subsubsection*{Summarization.} We give an overview of graph summarization methods, and refer the interested reader to a comprehensive survey~\cite{liu2018graph}. Most summarization works fall into 3 categories: (1) aggregation-based which group nodes~\cite{navlakha2008graph} or edges~\cite{maccioni2016scalable}) 
into super-nodes/edges based on application-oriented criteria or existing clustering algorithms; (2) abstraction-based which remove less informative nodes or edges; and (3) compression-based~\cite{shah2015timecrunch} which aim to minimize the number of bits required to store the input graph. 
Summarization methods have a variety of goals, including query efficiency, pattern understanding, storage reduction, interactive visualization, and domain-specific feature selection.
The most relevant work is CoSum~\cite{zhu2016unsupervised}, which tackles entity resolution by aggregating nodes into supernodes based on their labels and structural similarity.
Unlike these methods, \method applies to \textit{any type} of graphs and generates summaries independent of nodes/edges in a \textit{latent} graph space. Moreover, it is general and not tailored to specific ML tasks.

\begin{table}[t!]
	\vspace{-0.01cm}
	\centering
	\caption{Qualitative comparison of \method to existing summarization and embedding methods. Does the method: handle heterogeneous graphs; 
		yield an output that is size-independent, but node-specific, and representations that are independent of node proximity;
		support inductive learning and scale well (i.e., it is subquatratic on the network size)?}
	\label{table_comp}
	\vspace{-0.2cm}
	\resizebox{\columnwidth}{!}{
		\footnotesize
		\setlength{\tabcolsep}{2.6pt} \begin{tabularx}{1.0\linewidth}{@{}l@{} cc cccc ccc@{}}
			\toprule
			\BB
			& {\footnotesize \sc Input} && \multicolumn{3}{c}{\footnotesize \sc Representations / Output} &&  \multicolumn{2}{c}{\footnotesize \sc Method} \\ \cline{2-2} \cline{4-6} \cline{8-9}
			\TT
			& {\footnotesize \bf Hetero-} && {\footnotesize \bf Size} & {\footnotesize \bf Node} & {\footnotesize \bf Proxim.} &&&  \\ 
			& {\footnotesize \bf geneity} && {\footnotesize \bf indep.} & {\footnotesize \bf specific} & {\footnotesize \bf indep.} && {\footnotesize \bf Scalable} & {\footnotesize \bf Induc.}  \\ 
			\midrule
			Aggregation~\cite{blondel2008fast} & \cmark  && \xmark  & \xmark & \xmark && \cmark & \xmark  \\
			Cosum~\cite{zhu2016unsupervised} & \xmark  && \xmark  & \xmark & \cmark && \xmark & \xmark \\
			AspEm~\cite{shi2018aspem} & \cmark & & \xmark & \cmark & \xmark && \cmark & \xmark\\
			metapath2vec~\cite{dong2017metapath2vec} & \cmark && \xmark & \cmark & \xmark && \cmark & \xmark\\
			n2vec~\cite{node2vec}, LINE~\cite{line} & \xmark && \xmark & \cmark & \xmark && \cmark & \xmark \\
			struc2vec~\cite{struc2vec} & \xmark && \xmark  & \cmark & \cmark && \xmark & \xmark \\
			DNGR~\cite{cao2016deep} & \xmark  && \xmark  & \cmark & \xmark && \xmark & \xmark \\
			GraphSAGE~\cite{hamilton2017inductive} & \cmark  && \xmark  & \cmark & \cmark && \cmark & \cmark \\
			\midrule
			\method & \cmark && \cmark & \cmark & \cmark  && \cmark & \cmark\\
			\bottomrule
		\end{tabularx}
	}
\end{table}
\vspace{-0.1cm}

\section{Conclusion} 
\label{sec_conc}
This work introduced the problem of \emph{latent network summarization} and described a general computational framework, \method to learn such space-efficient latent node summaries of the graph that are completely independent of the size of the network.
The output (size) of latent network summarization depends only on the complexity and heterogeneity of the network, and captures its key structural behavior. Compared to embedding methods, the latent summaries generated by our proposed method require  $80$-$2152\times$ \textbf{less} output storage space for graphs with millions of  edges, while achieving significant improvement in AUC and F1 score for the link prediction task.
Overall, the experiments demonstrate the effectiveness of \method for link prediction, anomaly and event detection, as well as its scalability and space efficiency.

\section*{Acknowledgements}
{
\small 
This material is based upon work supported by the National Science Foundation under Grant No. IIS 1845491, Army Young Investigator Award No. W911NF1810397, an Adobe Digital Experience research faculty award, and an Amazon faculty award. 
Any opinions, findings, and conclusions or recommendations expressed in this material are those of the author(s) and do not necessarily reflect the views of the National Science Foundation or other funding parties. The U.S. Government is authorized to reproduce and distribute reprints for Government purposes notwithstanding any copyright notation here on.
}

\balance

\bibliographystyle{ACM-Reference-Format}
\bibliography{paper}

\newpage
\appendix

\label{sec_supp}
{\center \textbf{\LARGE Supplementary Material on Reproducibility}}
\section{Complexity Analysis Proof}
\label{sec_complexity_proof}
\subsection{Computational complexity}
\label{sec_computation_complexity}
\begin{proof}
	
	The computational complexity of \method includes deriving (a) distribution-based matrix representation $\mY$ and (b) its low-rank approximation.
	
	The computational unit of \method is the relational operation performed over the egonet of a specific node. Searching the neighbors for all node $i\in V$ has complexity $\mathcal{O}(N+M)$ through BFS. The complexity of step (a) is linear to $|\Fr|$,
	as indicated in \S~\ref{sec_summarizing_hetero}, this number is $|\mathcal{B}||\Phi|^\ell\cdot2|\mathcal{T}_V||\mathcal{T}_E|c$. 
	
	Based on the sparsity of $\mY$, \method performs SVD efficiently through fast Monte-Carlo Algorithm by extracting the most significant $K$ singular values~\cite{frieze2004fast} with computational complexity $\mathcal{O}(K^2 N)$.
	Therefore step (b) can be accomplished in $\mathcal{O}((K^{(\ell)})^2 N)$ by extracting the most significant $K^{(\ell)}$ singular values at level $\ell$. Furthermore, deriving all $K$ singular values has $\mathcal{O}(K^2N)$ complexity as $\sum^{L}_{\ell=1}(K^{(\ell)})^2 \le (\sum^{L}_{\ell=1}K^{(\ell)})^2 = K^2$. 
	The overall computational complexity is thus $\mathcal{O}( N|\Fr||\mathcal{T}_E||\mathcal{T}_V|c + K^2N + M)$.
	Note that both $|\Phi|$ and $L$ are small constants in our proposed method (\eg, $|\Phi|=7$ and $L\le 2$). \method scales linearly with the number of nodes and edges ($N+M$) in $G$.
\end{proof}

\subsection{Space Complexity}
\label{sec_space_complexity}
\begin{proof}
	In the runtime at level $\ell$, \method stores $\mX^{(\ell-1)}$ to dervie $\mX^{(\ell)}$ and $\mY^{(\ell)}$, which take $\mathcal{O}(N|\Fr|)$ and $\mathcal{O}(c|\Fr||\mathcal{T}_V| |\mathcal{T}_E|N)$ space, respectively. SVD can be performed with $p \ll N$ sampled rows.
	For the output, storing the set of ordered compositions of relational functions in the summary requires space complexity $O(|\Fr|)$. For the set of matrices $\mathcal{S}$, we store $\mathbf{S}^{(\ell)}$ across all $L$ levels. As shown in the time complexity analysis, the number of binned features (columns) in $\mY$ over all levels is $2|\Fr||\mathcal{T}_V||\mathcal{T}_E|c$, which includes incorporating $|\mathcal{T}_V|$ object types with both in-/out- directionality and all edge types.
	The size of the output summarization matrices is thus $\mathcal{O}(K|\Fr||\mathcal{T}_V||\mathcal{T}_E|c)$, which is related to the graph heterogeneity and structure and independent of the network size $N, M$.
\end{proof}

\section{Experimental Details}

\subsection{Configuration}
\label{sec_exp_conf}

We run all experiments on Mac OS platform with 2.5GHz Intel Core i7 and 16GB memory. 
We configure the baselines as follows: we use 2nd-LINE to incorporate 2-order proximity in the graph; we run node2vec with grid searching over $p,q \in \{0.25, 0.50, 1, 2, 4\}$ as mentioned in ~\cite{node2vec} and report the best. 
For GraRep, we set $k=2$ to incorporate 2-step relational information. For DNGR, we follow the paper to set the random surfing probability $\alpha=0.98$ and use a 3-layer neural network model where the hidden layer has 1024 nodes.
For Metapath2vec, we retain the same settings (number of walks $=1000$, walk length $=100$) to generate walk paths and adopt a similar the meta-path ``Type 1-Type 2-Type 1'' as the ``A-P-A'' schema as suggested in the paper. 
For \method, although arbitrary relational functions can be used, we use order-$1$ $f_b=\sum$ as the base graph function for simplicity in our experiments.
To begin with, we derive in-/out- and total degrees to construct the $N\times 3$ base feature matrix $\mX^{(0)}$ denoted as 
$\big[f_b\langle\vb_1, \neighborhood\rangle, f_b\langle\vb_2, \neighborhood\rangle, f_b\langle\vb_3, \neighborhood\rangle\big]$
\label{eq_base}
where $\vb_1=\mathbf{A}_{i:}$, $\vb_2=\mathbf{A}_{:i}$, and $\vb_3=(\mA+\mA^T)_{i:}$, for $i\in V$. 
We set $L=2$ to construct order-2 relational functions to equivalently incorporate 2-order proximity as LINE does, but we do not limit other methods to incorporate higher order proximity. All other settings are kept default. In table~\ref{table_operator}, we list all relational operators used in the experiment.

\renewcommand\arraystretch{0.7}
\begin{table}[h]
	\centering
	\vspace{-0.2cm}
	\caption{Relational operators used in the experiment } \vspace{-3mm}
	\centering 
	\fontsize{7}{7.5}\selectfont
	\setlength{\tabcolsep}{5pt} \label{table_def_operator}
	\def\arraystretch{1.6} \begin{tabularx}{0.95\linewidth}{!{\vrule width 0.6pt} Xl !{\vrule width 0.6pt} ll !{\vrule width 0.6pt}}
		\noalign{\hrule height 0.6pt}
		$\phi$ & \textbf{Definition} & $\phi$ & \textbf{Definition}
		\\
		\noalign{\hrule height 0.7pt}
		\textit{max/min} & $\max/\min_{i\in \mathcal{S}}\vx_i$ & \textit{variance} & $\frac{1}{|\mathcal{S}|}\sum_{i\in\mathcal{S}}\vx_i^2 - (\frac{1}{|\mathcal{S}|}\sum_{i\in \mathcal{S}}\vx_i)^2 $\\
		\textit{sum} & $\sum_{i\in \mathcal{S}}\vx_i$ & \textit{l1-distance} & $\sum_{j\in\mathcal{S}}|x_i-x_j|$\\
		\textit{mean} & $\frac{1}{|\mathcal{S}|}\sum_{i\in \mathcal{S}}\vx_i$ & \textit{l2-distance} & $\sum_{j\in\mathcal{S}}(x_i-x_j)^2$\\
		\noalign{\hrule height 0.7pt}
	\end{tabularx}
	\vspace{-0.3cm}
	\label{table_operator}
\end{table}

\subsection{Heterogeneous Context}
\label{sec_hetero_context_justification}

We justify the effectiveness to derive heterogeneous context of feature matrix $\mX$. We perform the link prediction task on yahoo-msg and dbpedia datasets w./w.o. using $\mY$ following the setup indicated in \S\ref{sec_exp_conf}. The result is as follows. We observe the significant improvement in performance when deriving the histogram heterogeneous context, which empirically supports our claim in \S\ref{sec_summarizing_hetero}.

\begin{table}[h]
	{\footnotesize
		\centering
		\caption{Link prediction performance w./w.o. $\mY$
		}
		\label{table_svd}
		\vspace{-2mm}
		\begin{tabular}{lr C{1.2cm} C{1.4cm}}
			\toprule
			Data & Metric & w.o. $\mY$ & w. $\mY$ 
			\\ \midrule
			yahoo-msg
			& \begin{tabular}{@{}r@{}}\textbf{AUC} \\ \textbf{ACC} \\\textbf{F1 macro}\end{tabular}
			& \begin{tabular}{@{}c@{}} 0.7919 \\ 0.7122 \\ 0.7111 \end{tabular}
			& \begin{tabular}{@{}c@{}} {\bf 0.8443} \\ {\bf 0.7587}  \\ {\bf 0.7577}\end{tabular} 
			\\\midrule
			dbpedia
			& \begin{tabular}{@{}r@{}}\textbf{AUC} \\ \textbf{ACC}  \\\textbf{F1 macro}\end{tabular}
			& \begin{tabular}{@{}c@{}} 0.9369 \\ 0.9023 \\ 0.9020 \end{tabular}
			& {\bf \begin{tabular}{@{}c@{}} 0.9820 \\ 0.9197  \\ 0.9197 \end{tabular}}
			\\
			\bottomrule
			
		\end{tabular}
		
		\vspace{-0.2cm}
	}
\end{table}

\subsection{Choice of initial vectors}
\label{sec_justify_init_features}

In this subsection we justify the impact of initial vectors $\mathcal{B}$ to the performance on two datasets, yahoo-msg and dbpedia. We follow the configuration in \S\ref{sec_exp_conf} and run \method with different initial vectors indicated in Table~\ref{table_init_vector}. We observe that there is no significant difference when using different $\mathcal{B}$. This is as expected as the three vectors are linearly correlated and shows the power of relational operators to derive complex higher-order features.
\begin{table}[h]
	{\footnotesize
		\centering
		\caption{Link prediction performance with different $\mathcal{B}$
		}
		\label{table_init_vector}
		\vspace{-2mm}
		\begin{tabular}{lr C{1.2cm} C{1.4cm} C{1.8cm}}
			\toprule
			Data & Metric & $\mathcal{B}=\{\vb_1\}$ & $\mathcal{B}=\{\vb_1, \vb_2\}$ & $\mathcal{B}=\{\vb_1, \vb_2, \vb_3\}$
			\\ \midrule
			yahoo-msg
			& \begin{tabular}{@{}r@{}}\textbf{AUC} \\ \textbf{ACC}  \\\textbf{F1 macro}\end{tabular}
			& \begin{tabular}{@{}c@{}} 0.8439 \\ 0.7558  \\ 0.7545 \end{tabular}
			& \begin{tabular}{@{}c@{}} {\bf 0.8443} \\ 0.7559 \\ 0.7547 \end{tabular}
			& \begin{tabular}{@{}c@{}} {\bf 0.8443} \\ {\bf 0.7587} \\ {\bf 0.7577}\end{tabular} 
			\\\midrule
			dbpedia
			& \begin{tabular}{@{}r@{}}\textbf{AUC} \\ \textbf{ACC}  \\\textbf{F1 macro}\end{tabular}
			& \begin{tabular}{@{}c@{}}\textbf{ 0.9821 } \\ 0.9164  \\ 0.9164 \end{tabular}
			& \begin{tabular}{@{}c@{}}\textbf{ 0.9821 }\\ 0.9168  \\ 0.9168 \end{tabular}
			& \begin{tabular}{@{}c@{}} 0.9820 \\ \textbf{ 0.9197 } \\ \textbf{ 0.9197 } \end{tabular}
			\\
			\bottomrule
			
		\end{tabular}
		
		\vspace{-0.2cm}
	}
\end{table}

\subsection{Choice of relational operators}

\label{sec_justification_operators}
In this subsection we justify the impact of relational operators $\mathbf{\Phi}$ to the performance on yahhoo-msg dataset. To explore the effect of each operator, we set $\mathcal{F}_b=\mathbf{\Phi}$, and set $\B=\{\mathbf{b}_1, \mathbf{b}_2, \mathbf{b}_3\}$ with $L=1$. To illustrate the effectiveness of the operators only, we do not derive  histogram-based node contexts. We observe that \emph{sum} and \emph{mean} have potentially higher impacts to the performance than \emph{max} and \emph{min}. Using the combination of \emph{l1-} and \emph{12-}distance produces the worst performance.

\begin{table}[h]
	{\footnotesize
		\centering
		\caption{Link prediction performance with different $\Phi$
		}
		\label{table_ablation}
		\vspace{-2mm}
		\begin{tabular}{l C{1.2cm} C{1.2cm} C{1.2cm}}
			\toprule
			$\mathbf{\Phi}$ & \textbf{AUC} & \textbf{ACC} & \textbf{F1 macro}
			\\ \midrule
			$\{\emph{max}\}$ & 0.7317 & 0.6582 & 0.6526 \\
			$\{\emph{max, min}\}$ & 0.7727 & 0.6888 & 0.6873 \\
			$\{\emph{sum, mean, var}\}$ & 0.8274 & 0.7584 & 0.7584 \\
			$\{\emph{l1, l2}\}$ & 0.6906 & 0.6311 & 0.6307 \\
			$\{\emph{max, min, sum}\}$ & 0.8283 & 0.7445 & 0.7436 \\
			$\{\emph{max, min, sum, mean}\}$ & {\bf 0.8336} & {\bf 0.7611} &  {\bf 0.7609} \\
			\bottomrule
			
		\end{tabular}
		
		\vspace{-0.1cm}
	}
\end{table}

\subsection{Detailed Event detection}
\label{sec_exp_enron}
Events detected in Fig.~\ref{fig_real_b}: (1)~The quarterly conference call where  Jeffrey Skilling, Enron's CEO, reports "outstanding" status of the company; (2)~The infamous quarterly conference call; (3)~FERC institutes price caps across the western United States; (4)~The California energy crisis ends; (5)~Skilling announces desire to resign to Kenneth Lay, founder of Enron; (6)~Baxter, former Enron vice chairman, commits suicide, and (7)~Enron executives Andrew Fastow and Michael Kopper invoke the Fifth Amendment before Congress.

\section{Heterogeneity context in detail}
\label{app_summarizing_hetero}

\subsection{Histogram representation}
Specific feature values in $\mX^{(\ell)}$ derived by $\ell$-order composed relational functions could be prodigious due to the power-law nature of real-world graphs (\eg, total degree), which leads to under-representation of other features in the summary.
Among various techniques to handle skewness, we describe the $N\times 1$ feature vector by the distribution of its unique values, on which we apply logarithmic binning~\cite{jin2017exploratory}.
The justification of our decision in Appendix~\ref{sec_hetero_context_justification} shows that binning is necessary to improve performance and incorporate multiple operators in \method.

For feature vector $\vx$, a set of nodes in $\neighborhood$ and $c$ bins, logarithmic binning returns a vector of length $c$:
\begin{equation}
\Psi(\vx, \neighborhood, c) = [C(0), C(1), \dots, C(\log_a{(c}))]
\label{eq_col_binning}
\end{equation}
where $C(v)$ counts the occurrence of value $v$: $C(v)=\sum_{i\in \neighborhood}\delta(v, \vx_i)$. In $C(v)$, $\delta$ is the Kronecker delta (a.k.a indicator) function, $a$ is the logarithm base, and $c=\max\{\max{(\vx)}, c\}$.
We set $c$ to the value exceeding the maximum feature value ($\max (\vx)$) regardless of object types to make sure that the output bin counts remain the same across all features. We can explicitly fill in 0s in Eq.~\eqref{eq_col_binning} in the case of $c>\max(\vx)$.
Similar to Eq.~\eqref{eq_base_vector}, we use $\Psi\langle\mathbf{x}, \mathcal{S'}, c\rangle$ to denote the process of applying $\Psi$ function over all nodes in $V$ (rows of $\mX$) to get the $N\times c$ log-based histogram matrix.
Further, we denote the process of applying $\Psi$ on all feature vectors (columns of $\mX$) as 
$\mY = \Psi\langle\mX, \neighborhood, c\rangle$. 
We use $\mY$ to denote the resultant histogram-based feature matrix.
In the next subsection, we will explain how to apply $\Psi$ on different related subsets $\mathcal{R}\subseteq\neighborhood$ to incorporate heterogeneity in the summary.

\subsection{Heterogeneous contexts}
\subsubsection{Object types} In heterogeneous graphs, the interaction patterns between a node and its typed neighbors reveal important behavioral information. 
Intuitively, similar entities have similar interaction patterns with every single type of neighbor.
For example, in the author-paper-venue networks, authors submitting papers to the same track at the same conference have higher similarity than authors submitting to different tracks at the same conference.
To describe how a specific node $i$ interacts with objects of type $t$, \method collects typed $t$ neighbors by setting $\mathcal{R}=\neighborhood^{t}_i$ and computes the ``localized'' histogram of a specific feature vector $\vx$ through $\Psi(\vx, \neighborhood^{t}_i, c)$. 
Repeating this process for nodes $i\in V$ forms an $N\times c$ distribution matrix $\Psi\langle \vx, \neighborhood^{t}, c\rangle$.

\begin{figure}[h]
	\centering
	\includegraphics[width=0.9\linewidth]{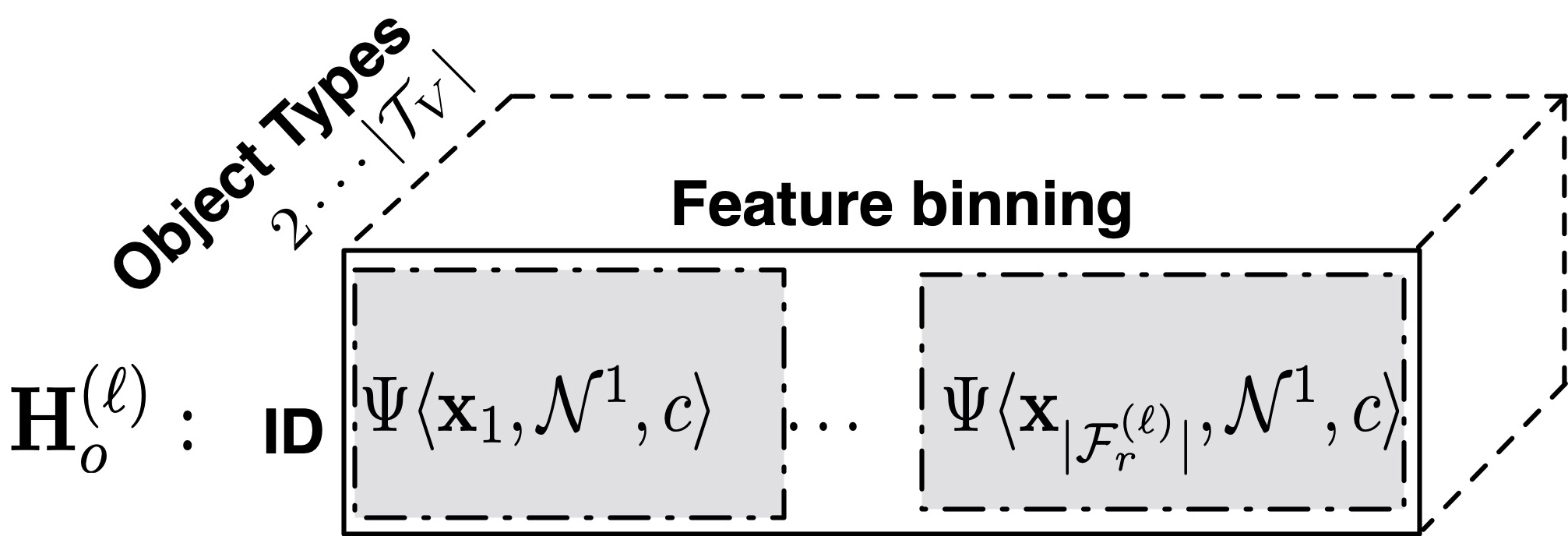}
	\vspace{-0.1cm}
	\caption{At level $\ell$, enumerating object types in $\neighborhood$ introduces one more dimension to the feature matrix and leads to a tensor. Note that $\Psi\langle\vx, \neighborhood, c\rangle$ is an $N\times c$ matrix (shaded area). Layer $t$ of the tensor is also denoted as $\Psi\langle\mX, \neighborhood^t, c\rangle$ for brevity.}         \label{fig_base_tensor_a}
	\vspace{-0.45 cm}
\end{figure}

\method enumerates all types of neighbors within $\neighborhood$ to incorporate complete interaction patterns for each node in the graph. This process can be seen as introducing one more dimension, the object types, to $\mY$ to form a tensor, as shown in Fig.~\ref{fig_base_tensor_a}.
We flatten the tensor through horizontal concatenation and denote it as $\mY$: \begin{equation}
\mY_{o}=\big[
\Psi\langle\mX, \neighborhood^{T_1}, c\rangle,
\Psi\langle\mX, \neighborhood^{T_2}, c\rangle,
\dots,
\Psi\langle\mX, \neighborhood^{t}, c\rangle
\big]
\label{eq_binning_ot}
\vspace{-0.2cm}
\end{equation}
where $t = \{1, \dots, |\mathcal{T}_{V}|\}$

\subsubsection{Edge directionality} So far we assume the input graph is undirected by focusing on nodes in $\neighborhood$ and search for neighbors in the 1-hop neighborhood regardless of edge directions. \method handles the directed input graphs by differentiating nodes from the out-neighborhood and in-neighborhood. 
The process is almost identical to the undirected case, but instead of using $\neighborhood$ in Equation~\eqref{eq_binning_ot}, we consider its two disjoint subsets $\neighborhood^{+}$ and $\neighborhood^{-}$ with incoming and outgoing edges, respectively.
The resultant histogram-based feature matrices are denoted as $\mY^{+}_{o}$ and $\mY^{-}_{o}$, respectively.
Again, we horizontally concatenate them to get the feature matrix incorporating edge directionality $\mY_{d}$ as $\mY_{d}=\big[
\mY_{o}^{+}, \mY_{o}^{-}
\big]$.

\subsubsection{Edge types}\label{sec_edge_types} Edge types in heterogeneous graphs play an important role in determining graph semantics and structure. The same connection between a pair of nodes with different edge types could convey entirely different meanings (e.g., an edge could indicate ``retweet'' or ``reply'' in a Twitter-communication network). 
This is especially important when the input is a multi-layer graph model.
To handle multiple edge types, \method constructs subgraphs $g(V, E_\tau)$ restricted to a specific edge type $\tau\in \mathcal{T}_E$.
For each subgraph, \method repeats the process to obtain the corresponding feature matrix $\mY_{d}$ per edge type that incorporates both node types and edge directionality.
We denote the feature matrix $\mY_{d}$ with respect to edge type $\tau$ as $\mY^{\tau}_{d}$. Thus, by concatenating them horizontally we obtain the final histogram-based context representation denoted as:
\begin{equation}
\mY_{e}=\big[
\mY_{d}^{1}, \mY_{d}^2, \dots, \mY_{d}^{\tau}
\big]
\label{eq_binning_et}
\end{equation}
where $\tau=\{1, \dots, |\mathcal{T}_E|\}$.
Therefore, $\mY_{e}$ captures all three aspects of heterogeneity of graph $G$ with size $N\times 2|\mathcal{T}_V||\mathcal{T}_E|c \cdot |\Fr|$. We use $\mY$ to denote this representation for brevity.

\end{document}